\documentclass[12pt,draftclsnofoot,peerreviewca,letterpaper,onecolumn]{IEEEtran}

\usepackage{cite}
\usepackage{graphicx,color,epsfig,rotating}
\usepackage{amsfonts,amsmath,amssymb,bbm}
\usepackage{algorithm, algorithmic}
\usepackage{amsmath}
\usepackage{cite}
\usepackage{mdwtab}
\usepackage{placeins}
\usepackage{psfrag, graphicx}
\usepackage[latin1]{inputenc}
\usepackage{amssymb}
\usepackage{makeidx}
\usepackage{epstopdf}
\usepackage{subcaption}
\usepackage{comment}
\usepackage{xcolor}

\newfont{\bbb}{msbm10 scaled 700}

\newfont{\bb}{msbm10 scaled 1100}

\newcommand{\Lc}{{\cal L}}
\newcommand{\Mc}{{\cal M}}

\newcommand{\eqdef}{\stackrel{\Delta}{=}}

\newcommand{\be}{\begin{equation}}
\newcommand{\ee}{\end{equation}}
\newcommand{\bea}{\begin{eqnarray}}
\newcommand{\eea}{\end{eqnarray}}

\newtheorem{defn}{Definition}
\newtheorem{example}{Example}
\newtheorem{theorem}{Theorem}
\newtheorem{lemma}{Lemma}
\newtheorem{corollary}{Corollary}
\newtheorem{remark}{Remark}

\begin{document}
\setcounter{page}{1}


\title{A Combinatorial Design for Cascaded Coded Distributed Computing on General Networks}

\author{Nicholas Woolsey,~\IEEEmembership{Student Member,~IEEE}, Rong-Rong Chen,~\IEEEmembership{Member,~IEEE},\\ and Mingyue Ji,~\IEEEmembership{Member,~IEEE} 
\thanks{This manuscript was partially presented in the conference papers \cite{woolsey2018new,woolsey2019cascaded}.}
\thanks{The authors are with the Department of Electrical Engineering,
University of Utah, Salt Lake City, UT 84112, USA. (e-mail: nicholas.woolsey@utah.edu, rchen@ece.utah.edu and mingyue.ji@utah.edu)}
}

\if{0}
\author{
    \IEEEauthorblockN{ Nicholas Woolsey,
		Rong-Rong Chen, and Mingyue Ji }
	\IEEEauthorblockA{Department of Electrical and Computer Engineering, University of Utah\\
		Salt Lake City, UT, USA\\
		Email: \{nicholas.woolsey@utah.edu,
		 rchen@ece.utah.edu,
		mingyue.ji@utah.edu\}}

}
\fi

\maketitle

\vspace{-0.5cm}

\begin{abstract}
 Coding theoretic approached have been developed to significantly reduce the communication load in modern distributed computing system. In particular,
coded distributed computing (CDC) introduced by Li {\em et al.} 
can efficiently trade computation resources to reduce the communication load in 
MapReduce like computing systems. For the more general \textit{cascaded} CDC, Map computations are repeated at $r$ nodes to significantly reduce the communication load among nodes tasked with computing $Q$ Reduce functions $s$ times. 
In this paper, we 
propose a novel low-complexity combinatorial design for cascaded CDC which 1) determines both input file and output function assignments, 2) requires significantly less number of input files and output functions, and 3) operates on heterogeneous networks where nodes have varying storage and computing capabilities.
We provide an analytical characterization of the computation-communication tradeoff, from which we show the proposed scheme can outperform the state-of-the-art scheme proposed by  Li {\it et al.} for the homogeneous networks. Further, when the network is heterogeneous, we show that the performance of the proposed scheme can be better than its homogeneous counterpart. In addition, the proposed scheme is optimal within a constant factor of the information theoretic converse bound while fixing the input file and the output function assignments.
\end{abstract}



\section{Introduction}
\label{section: intro}
Coded distributed computing (CDC), introduced in \cite{li2018fundamental}, provides an efficient approach to reduce the communication load by increasing the computation load in CDC networks such as MapReduce \cite{dean2008mapreduce} and Spark \cite{zaharia2010spark}. 
In this type of distributed computing network, in order to compute the output functions, the computation is decomposed into ``Map" and ``Reduce" phases. First, each computing node computes intermediate values (IVs) using  local input data files according to the designed Map functions. Then, computed IVs are exchanged among  computing nodes 
and nodes use these IVs as input to the designed Reduce functions to compute output functions.
The operation of exchanging IVs is called ``data shuffling" and occurs during the ``Shuffle" phase. This severely limits the performance of distributed computing applications due to the very high transmitted traffic load \cite{li2018fundamental}.

In \cite{li2018fundamental}, by formulating and characterizing a fundamental tradeoff between ``computation load" in the Map phase and ``communication load" in the Shuffle phase, Li  {\em et al.} demonstrated that these two quantities are approximately inversely proportional to each other. This means that if each IV is computed at $r$ carefully chosen nodes,
then the communication load in the Shuffle phase can be reduced by a factor of $r$ approximately. CDC achieves this multiplicative gain in the Shuffle phase by leveraging
coding opportunities created in the Map phase and strategically placing the input files among the computing nodes.
This idea was expanded on in \cite{konstantinidis2018leveraging,woolsey2018new} where new CDC schemes were developed. However, a major limitation of these schemes is that they can only accommodate homogeneous computing networks, i.e., the computing nodes have the same storage, computing and communication capabilities.

Understanding the performance potential and finding achievable designs for heterogeneous networks remains an open problem.
The authors in \cite{kiamari2017Globecom} derived a lower bound for the communication load for a CDC network where nodes have varying storage or computing capabilities. The proposed design achieves the optimum 
communication load for a system of $3$ nodes. 
In \cite{shakya2018distributed}, the authors studied CDC networks with $2$ and $3$ computing nodes where nodes have varying communication load constraints to find a lower bound on the minimum computation load. In our recent work \cite{woolsey2020cdc}, we proposed a new combinatorial design called {\em hypercuboid} for general heterogeneous CDC, where all the parameters can be arbitrarily large with some certain relationship due to the combinatorial nature of the design. The achievable communication load is optimal within a constant factor given the input file and Reduce function assignments.

In this paper, we focus on a specific type of CDC, called \textit{cascaded} CDC, where Reduce functions are computed at multiple nodes as opposed to just one node. According to our knowledge, other than \cite{li2018fundamental} and \cite{woolsey2018new}, the research efforts in CDC, including the aforementioned works, have focused on the case where each Reduce function is computed at exactly $s=1$ one node. However, in practice, 
it is often desired to compute each Reduce function 
$s>1$ times. This allows for consecutive Map-Reduce procedures as the Reduce function outputs can act as the input files for the next Map-Reduce procedure \cite{zaharia2010spark}. 
Cascaded CDC schemes of \cite{li2018fundamental} and \cite{woolsey2018new} are designed  to trade computing load for communication load.
However, the achievable schemes only apply to homogeneous networks. In addition, another major limitation for the original cascaded CDC design \cite{li2018fundamental} is the requirement of large numbers of both input files and reduce functions in order to obtain the promised multiplicative gain in terms of the communication load. 

{\em Contributions:} In this paper, first, we propose a novel combinatorial design for cascaded CDC on both homogeneous and heterogeneous networks where nodes have varying storage and computing capabilities. 
In particular, we show that the hypercuboid combinatorial structure proposed in \cite{woolsey2020cdc} can be applied for cascaded CDC in a non-straightforward way.
Meanwhile, the resulting computation-communication tradeoff achieves the optimal tradeoff within a constant factor 
given the input file and Reduce functions assignments.
Second, somehow surprisingly, compared to \cite{li2018fundamental}, the proposed design can achieve a better performance in terms of communication load not only in a heterogeneous network, but also in a homogeneous network while fixing other system parameters. We find the fundamental tradeoff proposed in \cite{li2018fundamental} is ``breakable" given the flexibility the proposed output function assignment (see the detailed discussion in Section~\ref{sec: Discussion}). In addition, in the heterogeneous network scenario, the proposed scheme can also outperform its homogeneous counterpart.
Third, the proposed design also greatly reduces the need for performing random linear combinations over IVs 
and hence, reduces the complexity of encoding and decoding in the Shuffle phase. Finally, the proposed design achieves an exponentially smaller required numbers of both input files and reduce functions in terms of the number of computing nodes.
To the best of our knowledge, this is the first work to explore heterogeneous cascaded CDC networks where Reduce functions are computed at multiple nodes. It offers the first general 
design architecture for heterogeneous CDC networks with a large number of computing nodes.

While the fundamentals of the hypercuboid  combinatorial framework  were first developed in \cite{woolsey2020cdc}, this work makes new contributions beyond those of \cite{woolsey2020cdc} in the following aspects:
\begin{itemize}
\item In this work, we extend the combinatorial framework of  \cite{woolsey2020cdc} to the more general setting of cascaded CDC, in which each reduced function is computed $r=s$ times across nodes in the network. In comparison, the design of   \cite{woolsey2020cdc} is for   $s=1$ only but with arbitrary $r$.
\item This work addresses new challenges in cascaded CDC including function assignments. To the best of our knowledge, this work is the first to develop a combinatorial design for cascaded function assignments for both homogeneous and heterogeneous networks. The combinatorial design of \cite{woolsey2020cdc} primarily focuses on input file mapping and IV shuffle method.
\item 
This work  develops a new multi-round Shuffle phase to meet the requirements of computing each reduced function $s$ times at multiple nodes. This multi-round Shuffle design, consisting of two shuffle methods, different from  that of \cite{woolsey2020cdc}, is applied to multiple rounds of IV shuffling to take advantage of the same set of IVs being requested at multiple nodes. This design is unique to the setting of cascaded CDC and is critical to minimize the communication load of the cascaded network.
The Shuffle phase in \cite{woolsey2020cdc}  is  single-round only due to the assumption of $s=1$.
\item  This work  shows that  the proposed  design using the
cascaded function assignment
can  break the fundamental limits presented in  \cite{li2018fundamental} not only in  heterogeneous networks, but also in  homogeneous networks.
Similar observation was made in  \cite{woolsey2020cdc} only for a heterogeneous network when $s=1$.
\end{itemize}

This paper is organized as follows. In Section \ref{sec: Network Model and Problem Formulation}, we present the network model and problem formulation. Then, we present the general scheme of the proposed cascaded CDC design in Section \ref{sec: seqd} and present design examples. 
In Section \ref{sec:optmality_load}, we present the achievable communication load and the optimality of the proposed design. 
In Section \ref{sec: Discussion}, we discuss the proposed scheme and compared its performance to 
the state-of-the-art design of \cite{li2018fundamental}. 
This paper will be concluded in Section \ref{sec: Conclusion}. All the proofs will be given in appendices.

\paragraph*{Notation Convention}
We use $|\cdot|$ to represent the cardinality of a set or the length of a vector. 
Also $[n] := [1,2,\ldots,n]$ for some $n\in\mathbb{Z}^+$, where $\mathbb{Z}^+$ is the set of all positive integers, and $\oplus$ represents bit-wise XOR.

\section{Network Model and Problem Formulation}
\label{sec: Network Model and Problem Formulation}

We consider a distributed computing network where a set of $K$ nodes, labeled as $[K]=\{1, \ldots , K \}$, have the goal of computing $Q$ output functions and computing each function requires access to all $N$ input files. The  input files, denoted $\{w_1 , \ldots , w_N \}$, have equal sizes with $B$ bits each. 
The set of $Q$ output functions is denoted by $\{ \phi_1 , \ldots \phi_Q\}$. Each node $k\in [K] $ is assigned to compute a subset of output functions, denoted by $\mathcal{W}_k \subseteq [Q] $. The result of output function $i \in [Q]$ is $u_i = \phi_i \left( w_1, \ldots , w_N \right)$.
Further, an output function can be computed  using ``Map" and ``Reduce" functions such that
$u_i = h_i \left( g_{i,1}\left( w_1 \right), \ldots , g_{i,N}\left( w_N \right) \right)$,
 where for each output function $i$ there exists a set of $N$ Map functions $g_{i,j}(w_j), j \in [N]$  
 and one Reduce function $h_i$. Furthermore, 
 we call the output of the Map function, $v_{i,j}=g_{i,j}\left( w_j \right)$, as the intermediate value resulting from performing the Map function for output function $i$ on file $w_j$. 
It can be seen that there are $QN$ intermediate values with $T$ bits each. 
Let each node have access to $M$ out of the $N$ files and 
let the set of files available to node $k$ be $\mathcal{M}_k \subseteq \{ w_1, \ldots , w_N\}$. 
The nodes use the Map functions to compute each intermediate value in the {\em Map} phase at least once. Then, in the {\em Shuffle} phase, nodes multicast the computed intermediate values among one another via a shared link 
so that each node can receive the necessary intermediate values that it could not compute itself. Finally, in the {\em Reduce} phase, nodes use the Reduce functions with the appropriate intermediate values as inputs to compute the assigned output functions. 

In this paper, we let each computing node computes all possible intermediate values from locally available files. 
Then, we 
let each of the $Q$ Reduce functions is 
computed at $s>1$ nodes 
where 
$s$ is the number of nodes which calculate each Reduce function. This scenario is called {\it cascaded} distributed computing  \cite{li2018fundamental} and is motivated by the fact that distributed computing systems generally perform multiple iterations of MapReduce computations. The results from the $Q$ output functions become the input files for the next iteration. To have consecutive Map Reduce algorithms which take advantage of the CDC, 
it is important that each output function is computed at multiple nodes. In addition,
we consider the general scenario where each computing node can have heterogeneous storage space and computing rescource. Our schemes accommodate heterogeneous networks in that nodes can be assigned a varying number of files and functions. 

The design of CDC networks
yields two important  parameters: the computation load $r$ and the communication load $L$. Here, $r$ is defined as the number of times each IV is computed among all computing nodes, or $r = \frac{1}{N}\sum_{k=1}^K|\Mc_k|$. In other words, $r$ is the number of IVs computed in the Map phase normalized by the total number of unique IVs, $QN$. The communication load $L$ is defined as the amount of traffic load (in bits) among all the nodes in the Shuffle phase 
normalized by $QNT$.
\begin{defn}
The optimal communication load is defined as 
\be
L^*(r,s) \eqdef \inf\{L: (r, s, L) \text{ is feasible}\}.
\ee
\end{defn}

\section{Hypercuboid Approach for Cascaded CDC}
\label{sec: seqd}
 In this section, we present the proposed  combinatorial design for general cascaded CDC networks that apply to both heterogeneous and homogeneous networks. We will begin with a simpler, two-dimensional example to introduce the basic ideas of the proposed approach. This is followed by a  description of the general scheme that includes four key components: Generalized Node Grouping, Node Group Mapping, Cascaded Function Mapping, and Multi-round Shuffle Phase.
We then present two three-dimensional examples of the proposed hypercuboid design, one for a homogeneous network, and one for a heterogeneous network, to further illustrate details of the proposed design and compute the achievable communication rates. 

\subsection{$2$-Dimensional Homogeneous Example}

\begin{figure}
\includegraphics[width=4cm]{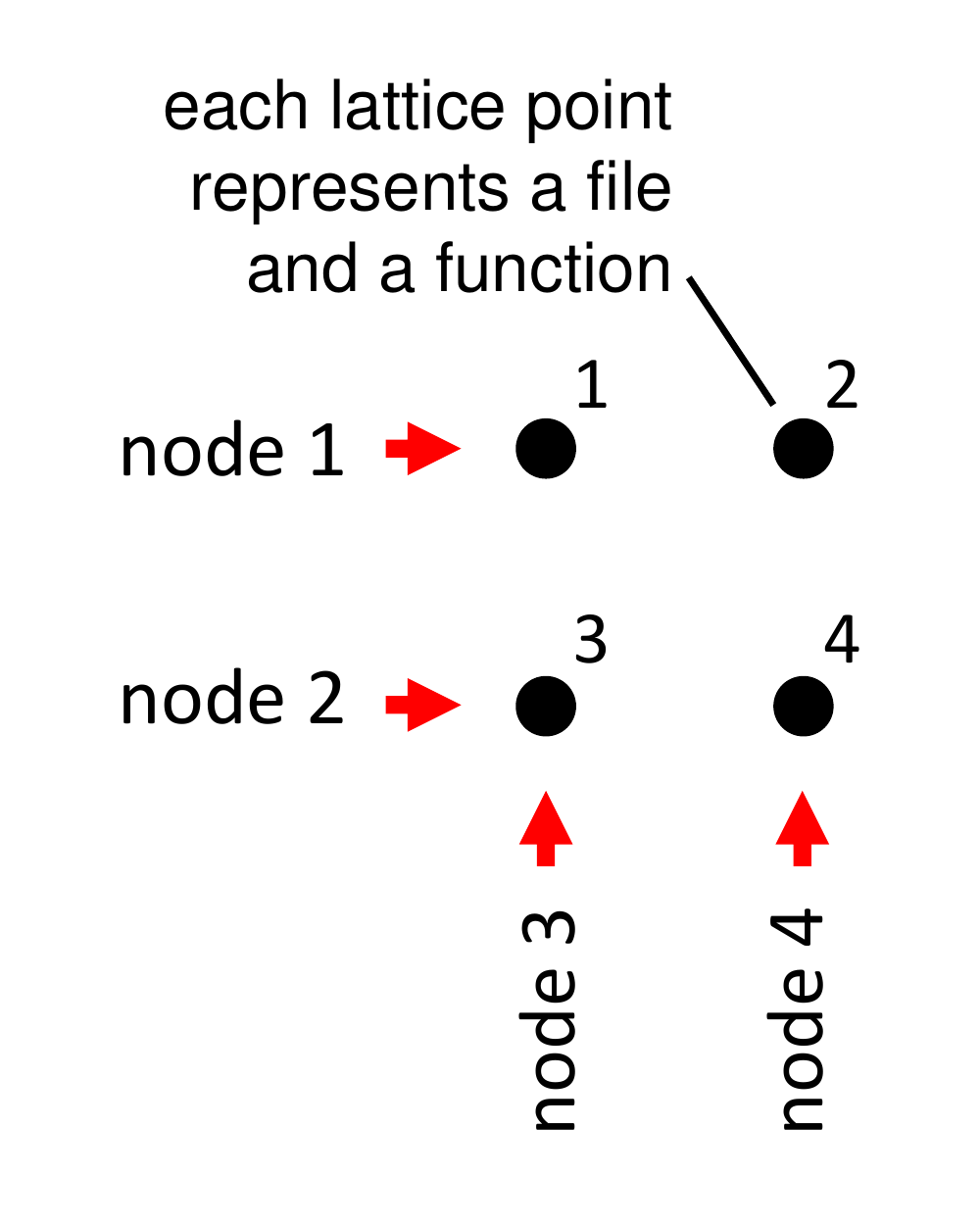}
\put(-2,27){\colorbox{gray!0}{\footnotesize
\begin{tabular}{|c|c|c|}
\hline
node & round $1$ & round $2$  \\
\hline
$1$ & $v_{1,2}\oplus v_{2,1}$ & $v_{3,2}^{(1)}+ v_{3,2}^{(2)}+ v_{4,1}^{(1)}+ v_{4,1}^{(2)}$  \\
 &  & $v_{3,2}^{(1)}-v_{3,2}^{(2)}-v_{4,1}^{(1)}+ v_{4,1}^{(2)}$  \\
\hline
$2$ & $v_{3,4}\oplus v_{4,3}$ & $v_{1,4}^{(1)}+ v_{1,4}^{(2)}+ v_{2,3}^{(1)}+ v_{2,3}^{(2)} $  \\
 &  & $v_{1,4}^{(1)}-v_{1,4}^{(2)} -v_{2,3}^{(1)}+ v_{2,3}^{(2)} $  \\
\hline
$3$ & $v_{1,3}\oplus v_{3,1}$ & $v_{2,3}^{(1)}+ v_{2,3}^{(3)}+ v_{4,1}^{(1)}+ v_{4,1}^{(3)} $  \\
 &  & $v_{2,3}^{(1)}-v_{2,3}^{(3)}-v_{4,1}^{(1)}+ v_{4,1}^{(3)} $  \\
\hline
$4$ & $v_{2,4}\oplus v_{4,2}$ & $v_{1,4}^{(1)}+ v_{1,4}^{(3)}+ v_{3,2}^{(1)}+ v_{3,2}^{(3)} $  \\
 &  & $v_{1,4}^{(1)}-v_{1,4}^{(3)}-v_{3,2}^{(1)}+ v_{3,2}^{(3)} $  \\
\hline
\end{tabular}
}}
\put(66,27){\colorbox{gray!0}{\footnotesize  
$
\underbrace{\left[
{\begin{array}{cccccc}
1 & 1 & 0 & 1 & 1 & 0\\[-0.4em]
1 & -1 & 0 & -1 & 1 & 0\\[-0.4em]
0 & 0 & 0 & 1 & 0 & 1\\[-0.4em]
0 & 0 & 0 & 1 & 0 & -1\\[-0.4em]
1 & 0 & 1 & 0 & 0 & 0\\[-0.4em]
1 & 0 & -1 & 0 & 0 & 0
\end{array}}
\right]}_{A}
\left[
\begin{array}{c}
v_{1,4}^{(1)} \\[-0.4em]
v_{1,4}^{(2)} \\[-0.4em]
v_{1,4}^{(3)} \\[-0.4em]
v_{2,3}^{(1)} \\[-0.4em]
v_{2,3}^{(2)} \\[-0.4em]
v_{2,3}^{(3)}
\end{array}
\right]
$
}}
\vspace{-0.5cm}
\caption{~\small (left) Lattice points that represent the file mapping and function assignment of the hypercube design with $r=2$, $s=2$ and $K=4$. Each lattice point represents a file and a function and each node maps files and is assigned files based on a line of lattice points. (middle) A table showing all the transmitted messages of the nodes. Each node can recover all requested IVs with the transmitted messages and the locally computed IVs. (right) The linear combinations of packets received by node $1$ in round $2$ after cancelling out locally computed packets.}
\label{fig: 2d}
\vspace{-0.6cm}
\end{figure}

\begin{example}\label{example:2d-homo}
Consider $K=4$ nodes that map $N=4$ files and are assigned to compute $Q=4$ functions. Fig.~\ref{fig: 2d} shows the file mapping and functions assignment. The nodes are aligned along a $2$-by-$2$ lattice and then horizontal or vertical lines define the mapping and assignment at the nodes. For instance, node $1$ maps files $w_1$ and $w_2$ and is assigned functions $1$ and $2$ represented by the top horizontal line of lattice points. Similarly, node $3$ maps files $w_1$ and $w_3$ and is assigned functions $1$ and $3$ represented by the left vertical line of lattice points. As each lattice point intersects $2$ lines, one vertical and one horizontal, then we find each file is mapped at $2$ nodes, $r=2$, and each function is assigned to $2$ nodes, $s=2$.

In the Map phase each node computes all IVs from each locally available file. For example, node $1$ computes $v_{1,1}$,  $v_{2,1}$, $v_{3,1}$, and  $v_{4,1}$ from file $w_1$ and $v_{1,2}$,  $v_{2,2}$, $v_{3,2}$, and  $v_{4,2}$ from file $w_2$. The IVs can be classified by the number of nodes that request them in the Shuffle phase. For instance, IV $v_{1,1}$ is only needed by nodes $1$ and $3$, but these nodes computes this IV from the locally available file $w_1$. Therefore, we say $v_{1,1}$ is requested by $0$ nodes. Similarly, $v_{2,2}$, $v_{3,3}$, and $v_{4,4}$ are requested by $0$ nodes. Then, since nodes $1$ and $3$ are the only nodes that are assigned function $1$, we see that 
$v_{1,3}$ is only requested by one node (node $1$) because $w_3$ is available at node 3, but not at node 1. Similarly, $v_{1,2}$ is only requested by one node (node $3$) because $w_2$ is available at node 1, but not at node 3.
On the other hand,  $v_{1,4}$ is requested by $2$ nodes, nodes 1 and 3, because s neither node maps file $w_4$.

There are $2$ rounds In the Shuffle phase  where the nodes shuffle the IVs that are requested by $1$ and $2$ nodes, respectively. The messages transmitted by each node are shown in the table of Fig.~\ref{fig: 2d}. In  round $1$, each node  computes $2$ IVs that are included in a coded message to $2$ other nodes. For instance, node $1$  computes $v_{1,2}$ and $v_{2,1}$, where $v_{1,2}$ which is requested by node $3$ and is available at node $4$, and the opposite is true for $v_{2,1}$. Therefore, node $1$ transmits $v_{1,2}\oplus v_{2,1}$ to nodes $3$ and $4$. In this example, each node  transmits a coded message to serve $2$ independent requests of nodes aligned along the other dimension. 

In round $2$, we consider all IVs requested by $2$ nodes which are $v_{1,4}$, $v_{4,1}$, $v_{2,3}$ and $v_{3,2}$. These IVs are each available at the $2$ nodes that do not request them. Each IV is split into $3$ disjoint equally sized packets. For instance, for the IVs requested by node $1$, $v_{1,4}$ is split into $v_{1,4}^{(1)}$, $v_{1,4}^{(2)}$ and $v_{1,4}^{(3)}$ and $v_{2,3}$ is split into $v_{2,3}^{(1)}$, $v_{2,3}^{(2)}$ and $v_{2,3}^{(3)}$. Each node sends two linear combinations of its available packets. Accordingly, each node will receive a total of $6$ linear combinations to solve for the $6$ requested packets. The linear combinations is shown in the table of Fig.~\ref{fig: 2d}. We can see, for instance, after subtracting out available packets, node $1$ receives the linear combinations shown by the matrix-vector multiplication on the right side of Fig.~\ref{fig: 2d}. Since, the matrix $A$ is invertible, node $1$ can solve for its requested packets and therefore all requested IVs. The messages of round $2$ are deliberately designed so that the received messages at each node can be represented by a full rank matrix similar to matrix $A$ for node $1$. One can verify from the the table of Fig.~\ref{fig: 2d}, that each node can recover all requested packets from round $2$. 

We compute the communication load, $L_{\rm c}$, by counting all transmitted messages and considering their size. There are $4$ messages of size $T$ bits (size of a single IV), and $8$ messages of size $\frac{T}{3}$ bits. After normalizing by the total bits over all IVs, $QNT$, the communication load is \begin{align}
    L_{\rm c} = \frac{1}{QNT}\left( 4T + 8\frac{T}{3} \right) = \frac{1}{16}\left( 4 + \frac{8}{3} \right) \approx 0.417.
\end{align}
With an equivalent, $r$, $s$ and $K$, we can compare $L_{\rm c}$ to the fundamental bound and scheme of \cite{li2018fundamental} where the communication load is
\begin{align}
    L_1 = \frac{3{K \choose 3}{1 \choose r-1}{r \choose 3-s}}{r{K\choose r}{K\choose s}} + \frac{4{K \choose 4}{2 \choose r-1}{r \choose 4-s}}{r{K\choose r}{K\choose s}} = \frac{3\cdot 4 \cdot 1 \cdot 2}{72} +\frac{4\cdot 1 \cdot 2 \cdot 1}{72} = \frac{32}{72}\approx 0.444.
\end{align}
Ultimately, we find $L_{\rm c}<L_1$, and our new design has a reduced communication load.
\end{example}
\begin{remark}
  To the best of our knowledge, this is the first homogeneous example of CDC that has a communication load less than $L_1$. In \cite{li2018fundamental}, $L_1$ was shown to be the smallest achievable communication load given $r$, $s$ and $K$,
  under an implicit assumption of the  reduce function assignment that {\it every} set of $s$ nodes must have have a common reduce function. This assumption was made in the proof of \cite[Theorem 2]{li2018fundamental}. Note that
our proposed design does not impose such an assumption. 
  For instance,  
  neither node pairs $\{1,2\}$  nor  $\{3,4\}$ in Example 1 have a  shared assigned function.  Example 1 shows that the more general function assignment proposed in this work allows us to achieve a lower communication load that is less than $L_1$, even for homogeneous networks. 
Similar observations were made for a heterogeneous network with $s=1$ in \cite{xu2019cdc,woolsey2020icc}. 
\end{remark}

\subsection{General Achievable Scheme}
\label{sec: gen_sd}

Next, we present the proposed general achievable scheme and describe its four key components in detail.

{\bf  Generalized Node Grouping} 
The Generalized Node Grouping lays the foundation of the proposed  hypercuboid design. It consists of Single Node Grouping (equivalent to Node Grouping 2 in \cite{woolsey2020cdc}), and  Double Node Grouping. The  latter is specifically designed for the cascaded CDC networks considered in this work.

Consider a general network of $K$ nodes with varying storage capacity. To define a hypercuboid structure for this network, divide these $K$ nodes into $P$ disjoint sets,  $\mathcal{C}_1,\ldots ,\mathcal{C}_P$, each of size $|\mathcal{C}_p|$ and  $\sum_{p=1}^{P}|\mathcal{C}_p|=K$.  Nodes in the same set $\mathcal{C}_p$ have the same storage capacity and each stores $1/m_p$ of the entire file library.
Furthermore, assume that nodes in $\mathcal C_p$ map the library $r_p$ times so that $\sum_{p=1}^P r_p=r$. Apply the hypercube design in \cite{woolsey2020cdc} to each $\mathcal C_p$ by splitting nodes in $\mathcal{C}_p$ into $r_p$ disjoint subsets $\{\mathcal{K}_{i}, i \in \mathcal I_p\}$ of equal size $m_p$, where $|\mathcal{C}_p|=r_pm_p,\;$ and the index set $\mathcal I_p=\{i: n_{p-1}+1 \le i \le n_p\}$ and $n_j=\sum_{i=1}^j r_i, j \in [P]$.   The entire network is comprised of $r$ node sets, $\mathcal{K}_1 ,\ldots , \mathcal{K}_r$. Nodes in $\mathcal K_i, i \in [r]$ are aligned along the $i$-th dimension of the hypercuboid, and they collectively map the library exactly once.  

{\bf Single Node Grouping} 
Given a subset $\mathcal A \subset [r]$,  we say that  $\mathcal S \subset \mathcal K$ is an $(\mathcal A,1)$ node group if it contains exactly one node from each $\mathcal K_i$, i.e., $|\mathcal{S}\cap\mathcal{K}_i|=1$, for every $i \in \mathcal A$. In particular, consider 
 all possible $([r],1)$ node groups $\mathcal{T}_1,\ldots ,\mathcal{T}_{X}$ of size $r$ that each  contains a single node from every node set $\mathcal{K}_1 ,\ldots , \mathcal{K}_r$, here  $X=\prod_{i=1}^{r}|\mathcal{K}_i|=\prod_{p=1}^{P}m_p^{r_p}$.  Denote $\mathcal T_{j,i}=\mathcal T_j \cap \mathcal K_i, \;\forall j \in [X]$ and $\forall i \in [r]$, as the node in $\mathcal T_j$ that is chosen from $\mathcal K_i$.

{\bf Double Node Grouping} 
Given a subset $\mathcal A \subset [r]$,  we say that  $\mathcal S \subset \mathcal K$ is an $(\mathcal A,2)$ node group if it contains exactly two nodes from each $\mathcal K_i$, i.e., $|\mathcal{S}\cap\mathcal{K}_i|=2$, for every $i \in \mathcal A$. Hence, the size of an  $(\mathcal A,2)$ node group is $|\mathcal{S}|=2\cdot |\mathcal{A}|$.
Double Node Grouping is essential for the design of the Multi-round Shuffle phase.

{\bf Node Group (NG) File Mapping}: Given all  $([r],1)$ node groups  $\mathcal{T}_1,\ldots ,\mathcal{T}_{X}$, 
we split the $N$ files  into $X$ disjoint sets labeled as $\mathcal{B}_{1},\ldots,\mathcal{B}_{X}$. These file sets are of size $\eta_1\in \mathbb{Z}^+$ and $N=\eta_1 X$. Each file set $\mathcal{B}_{i}$ is only available  to every node in the node group  $\mathcal{T}_i$.
  It follows that if  node $k \in [K]$ belongs to a node group $\mathcal T_i$, then the file set
  $\mathcal B_i$ is available to this node.
  Hence, by considering all possible node groups $\mathcal T_i$ that node $k$ belongs to,  its available files, denoted by $\mathcal{M}_k$, is expressed as 
\be
\mathcal{M}_k:=\bigcup\limits_{i : k\in \mathcal{T}_i}\mathcal{B}_i.
\ee

Note that, since each file belongs to a unique file set $\mathcal B_i$ and is mapped to a unique set of $r$ nodes (in the node group $\mathcal T_i$), we must have $\frac{1}{N}\sum_{k=1}^K|\Mc_k|= \frac{N r}{N}=r$.

The function assignment is defined as follows:

{\bf Cascaded Function Assignment}: Given  all  $([r],1)$ node groups $\mathcal{T}_1,\ldots ,\mathcal{T}_{X}$, 
the $Q$ files are split into $X$ disjoint sets labeled as $\mathcal{D}_{1},\ldots,\mathcal{D}_{X}$ and file set $\mathcal{D}_{i}$ is assigned exclusively to nodes of set $\mathcal{T}_i$. These function sets are of size $\eta_1\in \mathbb{Z}^+$ and $Q=\eta_2 X$.  For $k \in [K]$, define
\be
\mathcal{W}_k :=\bigcup\limits_{i : k\in \mathcal{T}_i}\mathcal{D}_i
\label{eq:fun_assign_3}
\ee
as the set of functions assigned to node $k$.

\begin{remark}
  Note that the proposed Cascaded Function Assignment follows the same design principle as that of the NG File Mapping. As each file is mapped to $r$ nodes in the network, the proposed design ensures that each reduce function is also mapped to $r$ nodes. Thus,  we assume that $r=s$ in our design. The proposed  Cascaded Function Assignment serves as a building block for consecutive rounds of MapReduce iterations, where the reduce function outputs become the file inputs for the next iteration.
\end{remark}



{\bf Map Phase}: Each node $k\in [K]$ computes the set of IVs $\{v_{i,j} : i\in [Q], w_j \in \mathcal{M}_k \}$.

\if
{\bf Multi-round (MR) Shuffle Phase:} We consider a Multi-round Shuffle Phase with $r$ rounds where
the $\gamma$-th round of the Shuffle phase is performed with one of two methods described below. 
In the MR Shuffle phase, we use two Shuffle Methods, termed  Inter-group (IG) Shuffle Method, and Linear Combination (LR) Shuffle Method,  to exchange IVs in the $\gamma$-th round, where the $\gamma$-th round is defined as the round in which IVs requested by $\gamma$ nodes are exchanged. 
The IG Shuffle Method  is designed for $1\le \gamma \leq r-1$ and forms groups of $2\gamma$ nodes. A node outside of each node group multicasts coded pairs of IVs to this node group. For the LC Shuffle Method, nodes also form groups of $2\gamma$ nodes; however, nodes of this group multicast linear combinations of packets among one another. The LC Shuffle Method is designed only for the $r$-th round. \mj{[MJ: but later you said this method can only work for $r$-th round.]}
\fi

{\bf Multi-round (MR) Shuffle Phase:} 
We consider a Multi-round Shuffle Phase with $r$ rounds, where in each round we use one of two methods termed the Inter-group (IG) Shuffle Method and the Linear Combination (LC) Shuffle Method. 
In the $\gamma$-th round, the nodes exchange IVs requested by $\gamma$ nodes. 
The IG Shuffle Method  is designed for $1\le \gamma \leq r-1$ and forms groups of $2\gamma$ nodes. A node outside of each node group multicasts coded pairs of IVs to this node group. For the LC Shuffle Method, nodes also form groups of $2\gamma$ nodes; however, nodes of this group multicast linear combinations of packets among one another. The LC Shuffle Method is designed only for the $r$-th round.

{\bf Inter-group (IG) Shuffle Method} ($1\leq \gamma \leq r-1$):  Consider $\mathcal A\subset [r]$ such that $|\mathcal{A}| = \gamma$. For each $\mathcal A$, let $\mathcal S$  be a $(\mathcal A,2)$-node group with $|\mathcal{S}|=2\gamma$ and  $\mathcal S' \subset \mathcal S$ be a $(\mathcal A,1)$-node group with $|\mathcal{S'}|=\gamma$. Assume $\mathcal A^c=[r]\setminus \mathcal A$ and let $\mathcal{Y}$ be a $(\mathcal A^c,1)$-node group with $|\mathcal{Y}|=r-\gamma$. An arbitrary node in $\mathcal Y$ will multicast a summation of two sets of IVs, one for nodes in $\mathcal{S'}$ and one for nodes in $\mathcal{S} \setminus \mathcal{S'}$. To ensure that each node in $\mathcal{S'}$ (or $\mathcal{S} \setminus \mathcal{S'}$) can decode successfully from the multicast message, the set of IVs intended for nodes in $\mathcal{S'}$ (or $\mathcal{S} \setminus \mathcal{S'}$) must be available to nodes in $\mathcal{S} \setminus \mathcal{S'}$ (or  $\mathcal{S'}$ ). To determine these IVs, letting
 $\mathcal{T}_\alpha = \{\mathcal{S}\setminus \mathcal{S}'\}\cup \mathcal{Y}$ and  $\mathcal{T}_\ell =  \mathcal{S}'\cup\mathcal{Y}$, we define
 \be
\mathcal{V}_{\mathcal T_\ell}^{\mathcal{S}\setminus \mathcal{S}'}  := \{ v_{i,j} : i\in \mathcal{D}_\alpha, w_j \in \mathcal{B}_\ell \} \quad \text{ and }\;\;
\mathcal{V}_{\mathcal T_\alpha}^{\mathcal{S}'} = \{ v_{i,j} : i\in \mathcal{D}_\ell, w_j \in \mathcal{B}_\alpha \Big\}.
\ee
By the definition of the NG File Mapping, nodes in $\mathcal T_\ell$ have access to files in in $\mathcal B_\ell$. However,
since nodes in $\mathcal{S}\setminus \mathcal{S}'$  are not in $\mathcal T_\ell$, they do not have access to files in $\mathcal B_\ell$.
Thus, the set
$\mathcal{V}_{\mathcal T_\ell}^{\mathcal{S}\setminus \mathcal{S}'}$ contains IVs that are requested by nodes in $\mathcal{S}\setminus \mathcal{S}'$  and can be computed at every node in $\mathcal T_\ell$. Similarly,
 $\mathcal{V}_{\mathcal T_\alpha}^{\mathcal{S}'}$ contains IVs that are requested by nodes in
 $\mathcal{S}'$ and can be computed at every node in  $\mathcal T_\alpha$.
For all possible choices of $\mathcal A, \mathcal S, \mathcal S', \mathcal Y$, an arbitrary node in 
$\mathcal{Y}$ multicasts
\be
\mathcal{V}_{\mathcal T_\ell}^{\mathcal{S}\setminus \mathcal{S}'} \oplus \mathcal{V}_{\mathcal T_\alpha}^{\mathcal{S}'}
\label{eq:def_message_shffule2}
\ee
to the $2\gamma$ nodes in $\mathcal{S}$.

{\bf Linear Combination (LC) Shuffle Method}  ($\gamma= r$): This shuffle method is used for the $r$-th round only.
Let $\mathcal S$ 
denote a $([r],2)$-node group with $|\mathcal{S}|=2r$.  Each node $k \in \mathcal S$ will multicast  linear combinations of IVs to the other $2r-1$ nodes in $\mathcal{S}$. These IVs are defined as follows. 
Given $k \in \mathcal S$, let $\mathcal T_{\ell} $ denote a $([r],1)$-node group such that  $k \in \mathcal T_{\ell} \subset \mathcal{S} $. Let $\mathcal T_{\alpha_\ell}=\mathcal S \setminus \mathcal T_{\ell}$, which is also a  $([r],1)$-node group.
Define 
\be
\mathcal{V}_{\mathcal T_{\ell}}^{\mathcal T_{\alpha_{\ell}}}  = \{ v_{i,j} : i\in \mathcal{D}_{\alpha_\ell}, w_j \in \mathcal{B}_\ell \},
\label{eq:ivs_shuffle3}
\ee
which are IVs requested by the $r$ nodes in $\mathcal T_{\alpha_\ell}$ and are available at the $r$ nodes in $\mathcal T_{\ell}$.
We then split each $\mathcal{V}_{\mathcal T_{\ell}}^{\mathcal T_{\alpha_\ell}}$ into $2r - 1$ equal size, disjoint subsets\footnote{ If the number of IVs in $\mathcal{V}_{\mathcal T_{\ell}}^{\mathcal T_{\alpha_\ell}}$ is not divisible by $2r-1$, then the IVs can be split into packets similar to Example \ref{example:2d-homo}.} denoted by
$ \mathcal{V}_{\mathcal T_{\ell},1}^{\mathcal T_{\alpha_\ell}}, \ldots , \mathcal{V}_{\mathcal T_{\ell},2 r -1}^{\mathcal T_{\alpha_\ell}}$.
 Let $\mathcal L_k=\{\ell: k \in \mathcal T_{\ell} \subset \mathcal{S} \}$.
 Then node $k$
 multicasts $2^{(r - 1)}$ linear combinations of the IVs in
\be
\bigcup\limits_{\ell \in \mathcal{L}_k}  \bigcup\limits_{ i \in [2r - 1 ]} \mathcal{V}_{\mathcal T_{\ell},i}^{\mathcal T_{\alpha_\ell}}
\ee
to the other $2r-1$ nodes in $\mathcal S$.

{\bf Reduce Phase}: For all $k\in [K]$, node $k$ computes all output values $u_q$ such that $q\in \mathcal{W}_k$.

\begin{remark}
The two Shuffle methods each have their own advantages. 
With the IG Shuffle Method, as shown in (\ref{eq:def_message_shffule2}), a node outside a node group $\mathcal S$ transmits a coded message containing $2$ IVs, one intended for $\gamma$ nodes in $\mathcal S'$, and one for  $\gamma$ nodes in $\mathcal S$. Hence, each transmission serves  $2\gamma$ nodes.
Moreover, the IG Shuffle Method  does not require the use of linear combinations or packetization of the IVs, and the sets of IVs can simply be XOR'd together. However, the IG Shuffle Method cannot be used in $r$-th Shuffle round since the node set $\mathcal{Y}$ would be empty. With the LC Shuffle Method, the node group shuffles linear combinations among one another and $2\gamma - 1$ nodes are served with each transmission. {Then, after a node receives all the transmissions from other nodes of the node group, it can solve for all its requested IV packets.} {While the LC Shuffle Method can be generalized for any round $\gamma$, since we only use it for $\gamma=r$, its generalized form is not presented here.} 
\end{remark}
\begin{comment}
\begin{remark} 
  \mj{When $r=2$, 
  in Section \ref{sec: Discussion}, we will discuss a special homogeneous CDC design which uses IG Shuffle Method  in the first Shuffle round which, surprisingly, has a smaller communication load than 
  that proposed in \cite{li2018fundamental}.}
\end{remark}
\end{comment}

\subsection{3-dimension Homogeneous Example}

\begin{example}
\label{example:3d-het-1}
{To demonstrate the general scheme, we construct a computing network using a 3-dimensional hypercube (or a cube) as shown in Fig.~\ref{fig: 3d fig_sgt1}. 
First, we present the {\em NG File Mapping}.\footnote{While the File Mapping in this example is the same as that in \cite{woolsey2020cdc}, it is included here for completeness.} Each lattice point in the cube 
represents a different file $\mathcal B_i=w_i$, $i\in [27]$ where $\eta_1=1$. The network has 
$K=9$ nodes, partitioned into three sets: $\mathcal{K}_1 = \{1,2,3 \}$,  $\mathcal{K}_2 = \{4,5,6 \}$, and $\mathcal{K}_3 = \{7,8,9 \}$, aligned along each of the $r=3$ dimensions of the cube. For example, the three nodes in $\mathcal{K}_1 = \{1,2,3 \}$ are represented by three parallel planes, e.g., node 3 is represented by the green plane. 
For file mapping, each node is assigned all files indicated by the 9 lattice points on the corresponding plane. For instance,
  node $5$, represented by the red plane,  is assigned the file set $\mathcal{M}_5=\{ w_2, w_5, w_8, w_{11}, w_{14}, w_{17}, w_{20}, w_{23}, w_{26}\}$.  For each $i \in [3]$,  the size of $\mathcal K_i$ is $\frac{K}{r}=3$, which is the number of lattice points in the $i$-th dimension. Since the three nodes in each set  $\mathcal{K}_i$ are aligned along dimension $i$, they collectively stores the entire  library of 27 files.
}Nodes compute every IV from each locally available file and therefore $r=3$.




Next, we illustrate the {\em Cascaded Function Assignment}. The reduce functions are assigned to multiple nodes by the same process as the file mapping. The planes representing the reduce functions (and files) for $6$ of $9$ nodes are shown in Fig.~\ref{fig: 3d fig_sgt1}. Each lattice point represents $\eta_2=1$ function (in addition to a file).  
For instance,  node $2$ stores  files $\{w_i\}$ for all $i\in \{ 4, 5, 6, 13, 14, 15, 22, 23, 24 \}$ (purple plane), 
and is assigned to compute reduce functions for all $i$ in the same set. 
A depiction of these planes are shown in Fig.~\ref{fig: 3d fig_sgt1} (a) and (b).

\begin{figure}
\centering
\centering \includegraphics[width=7.5cm]{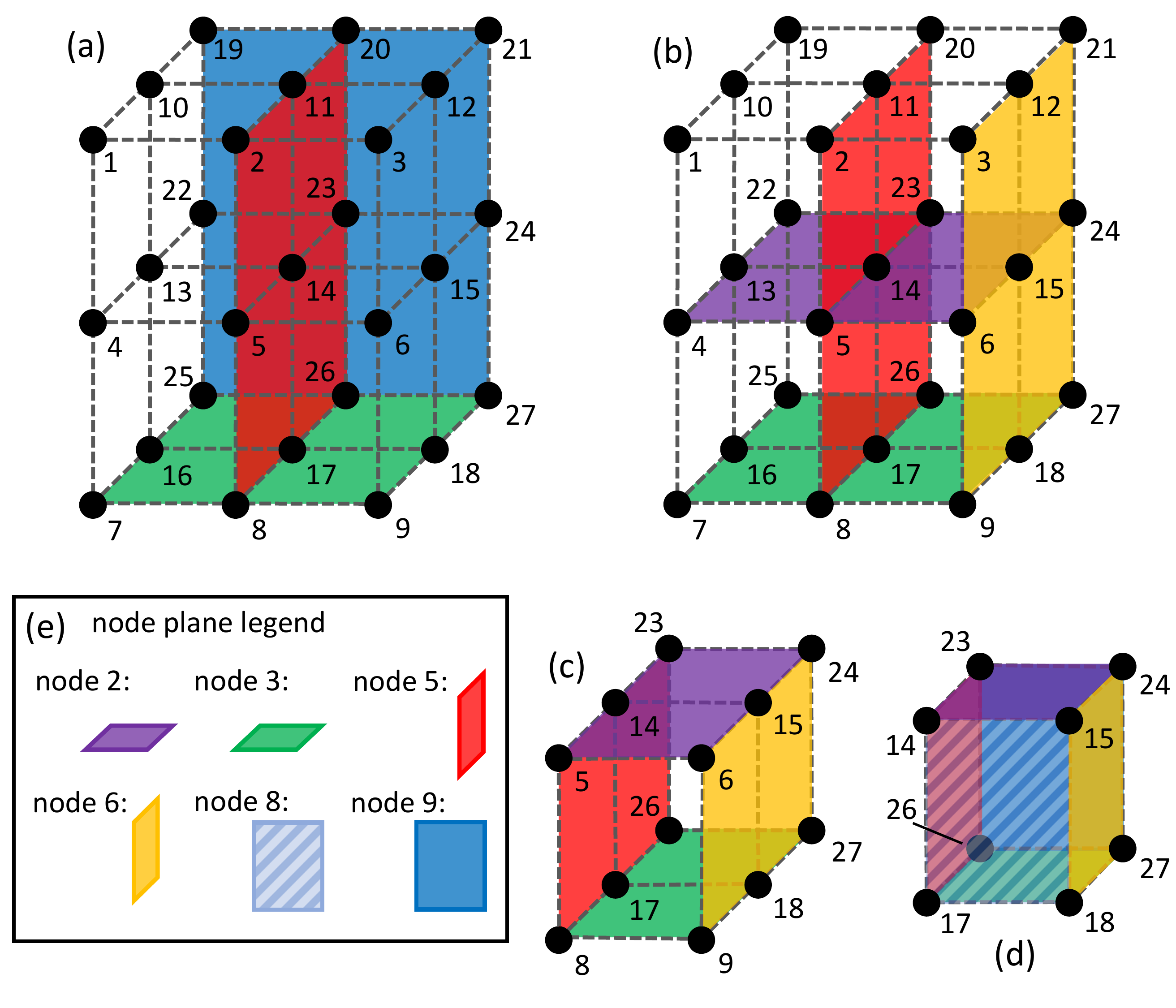} 
\vspace{-0.5cm}
\caption{~\small The input files and output functions  assigned to each node  are represented by planes of lattice points from a cubic lattice. (a) planes assigned to nodes of a set $\mathcal{T}_{26} = \{ 3,5,9 \}$. (b)  planes assigned to a set of nodes $\mathcal{S}=\{ 2,3,5,6\}$ in the 2nd round of the Shuffle phase and (c)  lattice points which intersect 2 planes of from nodes in $\mathcal{S}$. (d) the lattice points which intersect 3 planes of nodes in set $\mathcal{S} = \{ 2,3,5,6,8,9\}$. (e) legends for (a)-(d)}
\label{fig: 3d fig_sgt1}
\vspace{-0.4cm}
\end{figure}
 IVs can be categorized by the number ($\gamma$) of nodes which request them. 
Any IV of the form $v_{i,i}$ is only needed by nodes that can compute this IV themselves and are thus requested by $\gamma=0$ nodes. Next, we consider round $\gamma=1$ in which 
IVs which are requested by only $1$ node and use the IG Shuffle Method. These IVs can be identified by considering node group $\mathcal{T}_\alpha$ ($\alpha \in \{1, 2, \cdots, 27\}$), consisting of $1$ node from each set $\mathcal{K}_1$, $\mathcal{K}_2$ and $\mathcal{K}_3$. An example  is $\mathcal{T}_\alpha=\{3,5,9\}$ whose planes are depicted in Fig.~\ref{fig: 3d fig_sgt1}(a). Lattice points which fall on the intersection of exactly two of these planes represent input files that $2$ out of the $3$ nodes have available to it. As these $3$ nodes are the only nodes that compute the $26$-th reduce function, we 
see that $v_{26,23}$ is requested only by node $3$ and available at nodes $5$ and $9$. Next, consider another node group $\mathcal{T}_\ell=\{2,5,9\}$ that differs from $\mathcal{T}_\alpha$ by only in the first node (from $\mathcal{K}_1$). By observing the planes representing the nodes of $\mathcal{T}_\ell=\{2,5,9\}$, we find  $v_{23,26}$ is requested only by node $2$ and available at nodes $5$ and $9$. Therefore, either node $5$ or $9$ can transmit $v_{26,23}\oplus v_{23,26}$ to nodes $2$ and $3$ which can recover their requested IV. To match the description of the general scheme, we say $\mathcal{A}=\{1\}$, $\mathcal{S}=\{2,3 \}$,  $\mathcal{S}'=\{3 \}$ and $\mathcal{Y}=\{ 5,9 \}$. 

Next, we consider Shuffle round $\gamma=2$ in which 
IVs 
requested by $2$ nodes are exchanged using the IG Shuffle Method. Given $\mathcal{T}_{26} = \{3,5,9 \}$, whose planes are depicted in Fig.~\ref{fig: 3d fig_sgt1}(a), we consider lattice points which intersect only $1$ out of these $3$ planes. For instance, $w_{24}$ is available to node $9$ and not nodes $3$ or $5$. Therefore, nodes 3 and 5 are the only nodes that request IV $v_{26,24}$. Since node $9$ has this IV, it can multicast this IV to nodes $3$ and $5$. However, there is a way to serve two more nodes without increasing the communication load, recognizing that there are $2$ other nodes, $\{ 2,6 \}$,  that have input file $w_{24}$. 
Given $\mathcal{S} = \{2,3,5,6 \}$,  there is a set of 4 files $\{ w_{23}, w_{24}, w_{26}, w_{27} \}$ such that each file is only available to $2$ nodes in $\mathcal{S}$ and  all of these files are available to node 9. 
We define 
$\mathcal{S}'=\{3,5\}$ and $\mathcal{Y}=\{9\}$. Therefore, $ \mathcal{V}_{\{ 2,6,9 \}}^{\{ 3,5 \}} = \{ v_{26,24} \}$, $ \mathcal{V}_{\{ 3,5,9 \}}^{\{ 2,6 \}} = \{ v_{24,26} \} $ and node 9 transmits $v_{26,24} \oplus v_{24,26}$ to nodes $\{2,3,5,6 \}$. Keeping $\mathcal{Y}=\{ 9 \}$, we can also define $\mathcal{S}' = \{ 3,6\}$ to obtain $\mathcal{V}_{\{ 3,6,9 \}}^{\{ 2,5 \}} = \{v_{23,27}\}$ and $\mathcal{V}_{\{ 2,5,9 \}}^{\{ 3,6 \}} = \{v_{27,23}\}$ and node 9 also transmits $v_{23,27} \oplus v_{27,23}$ to nodes $\{2,3,5,6 \}$. Continuing with  $\mathcal{S}=\{ 2,3,5,6 \}$, consider lattice points which are in the planes parallel to plane of node $9$. These planes are defined by nodes $\{ 7,8 \}\in \mathcal{K}_3$. The lattice points of interests in regards to  $\mathcal{S}$ are highlighted in Fig.~\ref{fig: 3d fig_sgt1}(c).  We see that when $\mathcal{Y} = \{ 8 \}$, node $8$ transmits $ v_{17,15} \oplus  v_{15,17}$ and $v_{18,14} \oplus v_{14,18}$. When $\mathcal{Y} = \{ 7 \}$, node $7$ transmits $ v_{5,9} \oplus  v_{9,5}$ and $v_{6,8} \oplus v_{8,6}$.
Each node of $\mathcal{S}$ has locally computed one IV and requests the other IV from each of the transmissions from nodes $7$, $8$ and $9$.

Finally, we consider the last Shuffle round in which IVs requested by $\gamma=3$ nodes are exchanged by the LC Shuffle Method.
We see that none of the nodes in $\mathcal{T}_{26} = \{ 3,5,9 \}$ have access to  file $w_{15}$ and therefore, they all request $v_{26,15}$. All $3$ nodes  in  $\mathcal{T}_{15} = \{ 2,6,8 \}$  have computed $v_{26,15}$, but
request $v_{15,26}$ which  nodes of $\mathcal{T}_{26}$ have computed. In fact, any node in a $([\gamma],1)$ node group $\mathcal{S}' \subset \mathcal{S} = \mathcal{T}_{26} \cup \mathcal{T}_{15}$ computes an IV that nodes in $\mathcal{S} \setminus \mathcal{S}'$ request.
We consider the following sets of IVs: $\mathcal{V}_{\{ 2,6,8\}}^{\{ 3,5,9 \}} = \{ v_{26,15}\}$,
 $\mathcal{V}_{\{ 2,5,8\}}^{\{ 3,6,9 \}} = \{ v_{27,14}\}$,
  $\mathcal{V}_{\{ 3,6,8\}}^{\{ 2,5,9 \}} = \{ v_{23,18}\}$,
   $\mathcal{V}_{\{ 3,5,8\}}^{\{ 2,6,9 \}} = \{ v_{24,17}\}$,
   $\mathcal{V}_{\{ 3,5,9 \}}^{\{ 2,6,8\}} = \{ v_{15,26}\}$,
 $\mathcal{V}_{\{ 3,6,9 \}}^{\{ 2,5,8\}} = \{ v_{14,27}\}$,
  $\mathcal{V}_{\{ 2,5,9 \}}^{\{ 3,6,8\}} = \{ v_{18,23}\}$,
   and $\mathcal{V}_{\{ 2,6,9 \}}^{\{ 3,5,8\}} = \{ v_{17,24}\}$.
 The planes associated with each node of $\mathcal{S}=\{2,3,5,6,8,9 \}$ are highlighted in Fig.~\ref{fig: 3d fig_sgt1}(d).
 The IVs are split into $5$ packets so that each node requests $20$ unknown packets. Every node multicasts $4$ linear combinations of its computed packets so that every node receives transmissions from $5$ nodes and a total of $20$ linear combinations to solve for the $20$ requested packets. As proved in Appendix \ref{app:correctness}, at the end of $3$ shuffle rounds all node requests are satisfied.


In this example, each IV is computed at $3$ nodes and $r =  3$. 
 For round $1$ ($\gamma=1$),  there are $9$ node groups and nodes outside each group transmit an equivalent of $9$ IVs. This results in $9\cdot9=81$ transmissions. For round $2$ ($\gamma=2$), nodes form $27$ groups of $4$ nodes and nodes outside each group transmit an equivalent of $6$ IVs and leads to $27\cdot6=162$ transmissions.  For round 3 ($\gamma=3$), nodes form $27$ groups of $6$ nodes and each node in every group transmits an equivalent of $\frac{4}{5}$ IVs. This leads to $27\cdot6\cdot4/5=129.6$ transmissions of IVs. Collectively, the nodes transmit $(81+162+129.6)T=372.6T$ bits and thus
$L=\frac{372.6}{729} \approx 0.5111$.
\end{example}

\subsection{A 3-dimensional (cuboid) Heterogeneous Example}

\begin{example}
\label{example:3d-het-2}
Consider a heterogeneous network with $K=8$ computing nodes  where 
nodes $\{ 1,2,3,4\}$ have double the memory and computation power compared to nodes $\{ 5,6,7,8\}$.  Nodes are split into $3$ groups: $\mathcal{K}_1 = \{ 1,2\}$, $\mathcal{K}_2 = \{ 3,4\}$ and $\mathcal{K}_3 = \{ 5,6,7,8\}$. There are $X=16$ sets of functions and files and  node assignments are represented by a lattice structure (a cuboid) in Fig.~\ref{fig: het_exmp}. Let $\eta_1=\eta_2=1$ so that each  file set only contains $1$ file and each function set only contains $1$ function.
In the Map phase, every node computes every IV for each locally available file.

\begin{figure}
\centering \includegraphics[width=8cm]{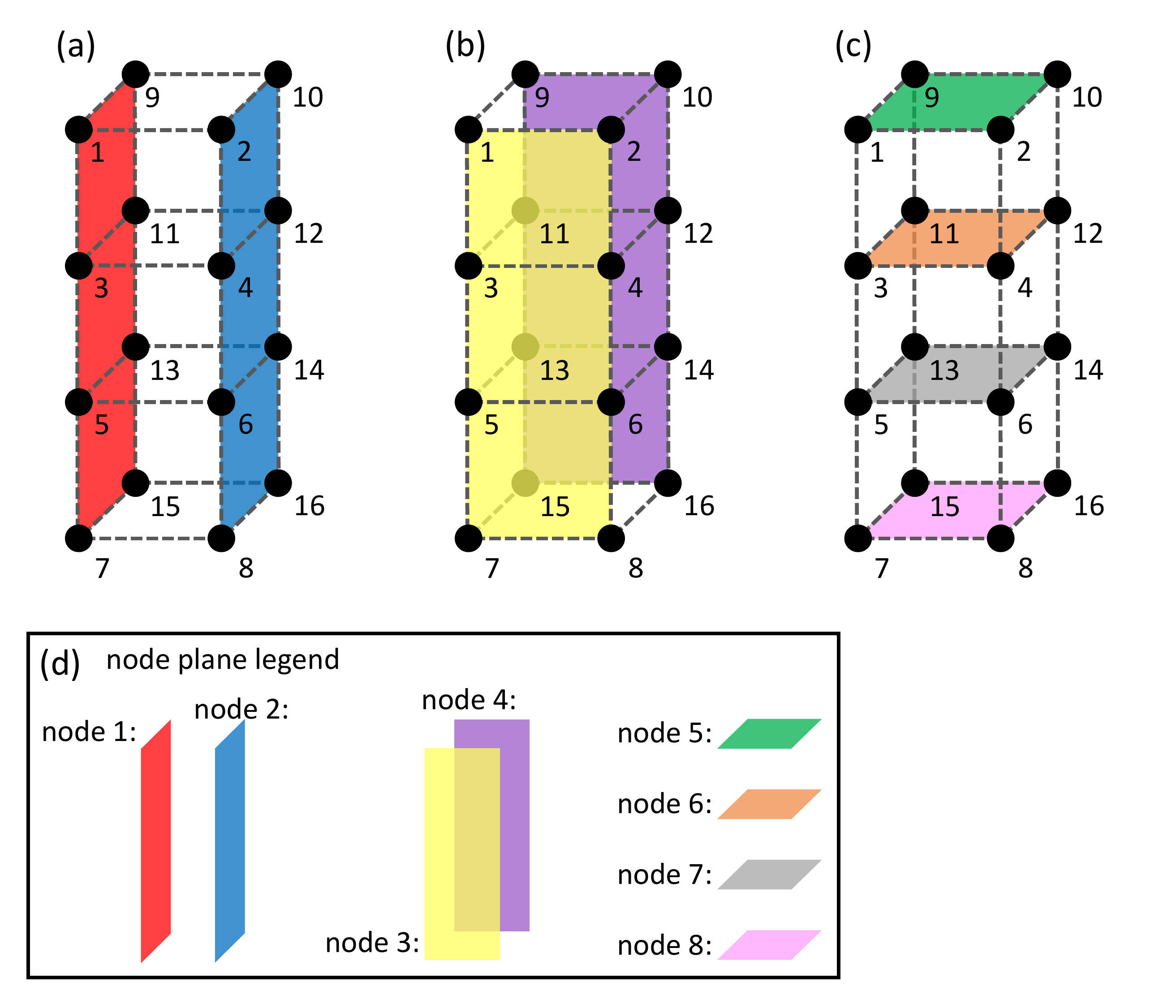} 
\vspace{-0.4cm}
\caption{~\small A 3-dimensional lattice that defines the file availability and reduce function assignment of 8 nodes in a heterogeneous CDC network. Each lattice point represents both a file and a function. Nodes are assigned files and functions represented by planes in the lattice.}
\label{fig: het_exmp}
\vspace{-0.4cm}
\end{figure}

Next, we consider the Shuffle phase. In round $1$ ($\gamma=1$), we use the IG Shuffle Method  and consider pairs of nodes that are from the same set $\mathcal{K}_i$ and aligned along the same dimension. Let $\mathcal{S} = \{ 1,2 \}, \mathcal{S'} = \{ 1\}$, and $\mathcal{Y}=\{3,8 \}$. We then have $\mathcal{S'} \cup \mathcal{Y}=\{1,3,8\}=\mathcal{T}_7$, 
and  $(\mathcal{S}\setminus \mathcal{S'}) \cup \mathcal{Y}=\{2,3,8\}=\mathcal{T}_8$.  Note that node $1$ is the only node that requests $v_{7,8}$ and node $2$ is the only node that requests $v_{8,7}$. Hence, either node $3$ or $8$ from $\mathcal{Y}$ can transmit $v_{7,8} \oplus v_{8,7}$ to nodes $1$ and $2$ in $\mathcal{S}$. 
Continuing this process, we see that all IVs requested by a single node are transmitted in coded pairs.

Next, for round $2$ ($\gamma=2$) we use the IG Shuffle Method  and consider groups of $4$ nodes where $2$ are from $\mathcal{K}_i$ and $2$ are from $\mathcal{K}_j$ where $i\neq j$. For instance, let $\mathcal{S}=\{3,4,6,8\}$.
 If we let $\mathcal{S'} = \{ 3,6\}$, and $\mathcal{Y}=\{1 \}$. We then have $\mathcal{S'} \cup \mathcal{Y}=\{3,6,1\}=\mathcal{T}_3$, and  $(\mathcal{S}~\setminus~ \mathcal{S'}~) ~\cup~ \mathcal{Y}~=~\{4,8,1\}=\mathcal{T}_{15}$.
 Thus, node $1$ from $\mathcal{Y}$ will transmit $v_{3,15}\oplus v_{15,3}$ to $\mathcal{S}$.
 For the same $\mathcal{S}=\{3,4,6,8\}, \mathcal{S'} = \{ 3,6\}$, if we let $\mathcal{Y}=\{2 \}$, then we have $\mathcal{S'} \cup \mathcal{Y}=\{3,6,2\}=\mathcal{T}_4$, and  $(\mathcal{S}\setminus \mathcal{S'}) \cup \mathcal{Y}=\{4,8,2\}=\mathcal{T}_{16}$.
  Thus, node 2 from $\mathcal{Y}$ will transmit $v_{4,16}\oplus v_{16,4}$ to $\mathcal{S}$. 
 Hence, IVs requested by $2$ nodes can also be transmitted in coded pairs.

Finally, for $3$ ($\gamma=3$) we use the LC Shuffle Method  and consider groups of $6$ nodes that contains $2$ nodes from each set $\mathcal{K}_1$, $\mathcal{K}_2$ and $\mathcal{K}_3$. For instance, consider $\mathcal{S} = \{ 1,2,3,4,5,6\}$. If we choose $\mathcal{S'} = \{ 1,3,5\} = \mathcal{T}_1$, then we have $\mathcal{S}\setminus \mathcal{S'}=\{2,4,6\}=\mathcal{T}_{12}$. We observe that $v_{1,12}$ is requested by three nodes in $\mathcal S'$ and is computed by all three nodes in   $\mathcal{S}\setminus \mathcal{S'}$. Similarly, we consider the other three cases: $\mathcal{S'} = \{ 1,3,6\} = \mathcal{T}_3$,
 $\mathcal{S}\setminus \mathcal{S'}=\{2,4,5\}=\mathcal{T}_{10}$; $\mathcal{S'} = \{ 1,4,5\} = \mathcal{T}_9$,
 $\mathcal{S}\setminus \mathcal{S'}=\{2,3,6\}=\mathcal{T}_{4}$; and $\mathcal{S'} = \{ 1,4,6\} = \mathcal{T}_{11}$,
 $\mathcal{S}\setminus \mathcal{S'}=\{2,3,5\}=\mathcal{T}_{2}$. In this way, we identify
 8 IVs which are requested by 3 nodes of $\mathcal{S}$ and locally computed at the $3$ other nodes of $\mathcal{S}$. These IVs are: $v_{1,12}$, $v_{12,1}$,  $v_{3,10}$, $v_{10,3}$, $v_{4,9}$, $v_{9,4}$, $v_{2,11}$ and $v_{11,2}$. Each IV is then split into $2 \gamma -1=2 \cdot 3 -1=5$ equal size packets and each node of $\mathcal{S}$ transmits $2^{\gamma-1}= 2^2=4$ linear combinations of its locally available packets. Each node collectively receives $4 \cdot 5=20$ linear combinations from the other $5$ nodes in $\mathcal S$ which are sufficient to solve for the requested $4$ IVs or $20$ unknown packets.

In this example, the computation load is $r=3$ because every file is assigned to $3$ nodes and every node locally computes all possible IVs.
In order to compute the communication load, we can see that IVs requested by $0$ nodes do not have to be transmitted. IVs requested by $1$ or $2$ nodes are transmitted in coded pairs, effectively reducing the communication load by half to shuffle these IVs. Hence, the number of transmissions in round $1$ and $2$ are given by  $\frac{80}{2}=40$, and $\frac{112}{2}=56$, respectively. The number of transmissions in round 3 is $6 \cdot 6 \cdot \frac{4}{5}= 28.8$ because there are $6$ choices of $\mathcal{S}$ of size $6$ and each node transmit effectively $\frac{4}{5}$ of an IV.
The communication load is thus given by  $L = \frac{40+56+28.8}{256}= 0.4875$ where $QN=16\cdot 16 =256$.
\end{example}

\section{Achievable  Communication Load and Optimality}\label{sec:optmality_load}
 In this section, we present the achievable communication load of the proposed design for general cascaded CDC networks and  discuss the optimality of the design given the proposed file and function assignment. An example is provided to illustrate the key steps in finding an information theoretic  lower bound on the achievable communication load. 
\subsection{Achievable communication load}


\begin{theorem}\label{theorem: sg1het}
For the proposed hypercuboid scheme with NG File Mapping, Cascaded Function Assignment, and Multi-round Shuffle Phase, 
the following communication load is achievable
\begin{eqnarray}
\label{eq: eqL_sgt1}
L_{\rm c} &=&  \frac{X-1}{2X} + \frac{1}{X(4r-2)}\prod_{i=1}^{r} (|\mathcal{K}_i|-1), 
\end{eqnarray}
where $X=\prod_{i=1}^{r}|\mathcal{K}_i|$.
 An upper bound on $L_c$ is obtained from  (\ref{eq: eqL_sgt1}) as
\be
L_{\rm c} < \frac{1}{2} + \frac{1}{4r-2}=\frac{r}{2r-1} \leq \frac{2}{3}.
\label{eq:L_c_upper}
\ee
\end{theorem}
\begin{IEEEproof}
The proof of Theorem \ref{theorem: sg1het} is given in Appendix \ref{sec: achieve sg1}.
\end{IEEEproof}
\begin{corollary}
When  setting $|\mathcal{K}_i|=\frac{K}{r}$ and $X=(\frac{K}{r})^r$, 
(\ref{eq: eqL_sgt1}) gives  the $L_c$ of a homogeneous network with parameters $K$ and $r$. 
\end{corollary}
\subsection{Optimality}
\label{sec: opt_exp}
In this section, we will show the optimality of the proposed hypercuboid approach for cascaded CDC. 
Note that, the fundamental computation-communication load tradeoff of \cite{li2018fundamental} does not apply to the cascaded CDC design since it has a different reduce function assignment compared to that of \cite{li2018fundamental}. We start by presenting the optimality of a homogeneous network using our proposed design and then for the more general heterogenous design.

\begin{theorem}\label{theorem: sg1_homo_bound}
Consider a homogeneous system with parameters $K$ and $r$.
Let $L^*$ be the infimum of achievable communication load over all possible shuffle designs given the proposed NG File Mapping and  Cascaded Function Assignment. Then, we have
\begin{align}
L^* &\geq 
\frac{1}{2}-\frac{1}{4m-2} - \left(\sum_{\hat{m}=1}^{m-1}\frac{ \hat{m}^{2r}}{4\hat{m}^2-1}\right)  m^{-2r}
\label{eq:bound_homo_L*}
\end{align}
where $m=\frac{K}{r}\geq 2$. Furthermore, given $L_c$ in (\ref{eq: eqL_sgt1}), it follows from (\ref{eq:bound_homo_L*}) that $L_c$ is within a constant multiple of $L^*$
\be
L_{\rm c} \leq \frac{64}{29}\cdot L^* \approx 2.207 L^*,
\label{eq:Lc_upper}
\ee
for general $K,r$. 
\end{theorem}
\begin{IEEEproof}
Theorem \ref{theorem: sg1_homo_bound} is proved in Appendix \ref{sec: lower bound}.
\end{IEEEproof}
\begin{remark}
 For the homogeneous network in Example 2, we have $L_c=0.5111$. This is compared to the lower bound of (\ref{eq:bound_homo_L*}) that gives $L^* \ge 0.3937. $ In this case, we have $L_c \le 1.2982 L^*$, achieving a better constant than that of the general case given in (\ref{eq:Lc_upper}).
\end{remark}

\begin{theorem}\label{th: conv}
Consider a general heterogeneous system with parameters $K$ and $r$.
Let $L^*$ be the infimum of achievable communication load over all possible shuffle designs given the proposed NG File Mapping and  Cascaded Function Assignment. Let $x_i =  |\mathcal{K}_i|$. Without loss of generality, assume that $x_1 \ge x_2 \cdots \ge x_s$. Then, we have
\begin{align}
L^* \ge \text{max}(L_{P1},L_{P2}),
\label{eq:two_LB}
\end{align}
where 
\be
\label{eq:opt_permutation}
L_{P1}=\frac{x_1-1}{2x_1};\quad 
L_{P2}=\frac{1}{X^2} \sum_{i=1}^r\Bigg( \prod_{j=1}^{i-1}x_j^2\Bigg)\sum_{\hat{m}=x_{i+1}+1}^{x_i}\sum_{\ell=1}^{i}(\hat{m}-1)^{2\ell-1}\hat{m}^{2(i-\ell)}.
\ee
Furthermore, for general $K$ and $r$, we show that $L_c$ is within a constant multiple of $L^*$,
\be
L_{\rm c} < \frac{8}{3}\cdot L^*.
\label{eq:constant_het}
\ee
\end{theorem}
\begin{IEEEproof}
Theorem \ref{th: conv} is proved in Appendix \ref{sec: het_opt_pf}.
\end{IEEEproof}

\begin{remark}
The two lower bounds $L_{P1}$ and $L_{P2}$  in (\ref{eq:opt_permutation}) correspond to two different choices of permutations used to evaluate the right side of (\ref{eq: bound_eq1}) of Lemma \ref{theorem: bound} in Appendix \ref{sec: lower bound}.
Extensive simulations suggest that the permutation used in $L_{P2}$ is optimal {in  achieving the largest lower bound using Lemma 2}. For  instance, consider Example 4, we get $L_{P1}=0.375$, which is less than $L_{P2}=0.3945.$ However, due to the complexity of (\ref{eq:opt_permutation}), we use the simpler $L_{P1}$ to determine the constant in (\ref{eq:constant_het}). Note that the permutation used for $L_{P2}$ matches with the permutation used in the homogeneous case to derive (\ref{eq:bound_homo_L*}).
Since $L_{P1}$ is in general weaker than $L_{P2}$, we see that the constant in (\ref{eq:constant_het}), derived using $L_{P1}$, is larger than that of (\ref{eq:Lc_upper}) for the homogeneous case. 
\end{remark}

\subsection{Optimality Example}
\begin{example}
\label{ex:optimality}
This example shows how to find a lower bound on the achievable communication load given the proposed NG Filing Mapping and Cascaded Function Assignment. Here, we use the homogeneous file mapping and function assignment of Example 2. Our approach builds upon an information theoretic lower bound (\ref{eq: bound_eq1}) 
(see  Lemma \ref{theorem: bound} and the notations therein in  Appendix \ref{sec: lower bound}),  originally designed for $s=1$ in \cite{woolsey2020cdc}, and extend it to the case of $s>1$ for the case of cascaded CDC.  

Lemma \ref{theorem: bound} requires that we pick a permutation of nodes and then use file and function counting arguments. 
The permutation we use here is $\{1,7,6,2,8,5,3,9,4\}$. To achieve a tighter bound, this permutation contains $3$ sequential node groups where each node group contains one node aligned along each dimension of the cube. In order to calculate the  terms of (\ref{eq: bound_eq1}), for each node $k_i$, we count the number of files not available to the first $i$ nodes of the permutation. This set of files is $\mathcal{M}_{\mathcal{K}}\setminus \mathcal{M}_{\{k_1,\ldots , k_{i} \}}$, called file of interests for node $k_i$. We also count the the number of functions assigned to the $i$-th node of the permutation that are not assigned to the previous $i-1$ nodes.  This set of functions is $\mathcal{W}_{k_i}\setminus \mathcal{W}_{\{k_1,\ldots , k_{i-1} \}}$, called functions of interests for node $k_i$. The product of these file and function counts represents the number of IVs of interests in Lemma  \ref{theorem: bound}. Moreover, since the IVs are independent and of size $T$ bits, we have
\begin{equation}
\label{eq:count}
H\left(V_{\mathcal{W}_{k_i},:}|V_{:,\mathcal{M}_{k_i}},Y_{\{k_1,\ldots, k_{i-1} \}}\right)  =T\cdot \left|\mathcal{M}_{\mathcal{K}}\setminus \mathcal{M}_{\{k_1,\ldots , k_{i} \}}\right| \cdot \left|\mathcal{W}_{k_i}\setminus \mathcal{W}_{\{k_1,\ldots , k_{i-1} \}}\right|.
\end{equation}
In Fig.~\ref{fig: 3d_proof_fig}, we  highlight the lattice points representing the sets of files and functions which are used to obtain the bound. Lattice points representing the files are highlighted in red and  lattice points representing the functions are highlighted in green. First, we consider every function assigned to node $1$ and every file not available to node $1$, where node $1$ is the first node in the permutation. This is shown in  Fig.~\ref{fig: 3d_proof_fig}(a). We see that
$H\left(V_{\mathcal{W}_{1},:}|V_{:,\mathcal{M}_{1}}\right) = T\cdot 18 \cdot 9 = 162T$
since in this case each lattice point only represents $1$ file and $1$ function ($\eta_1=\eta_2=1$).
\begin{figure}
\centering
\centering \includegraphics[width=8cm]{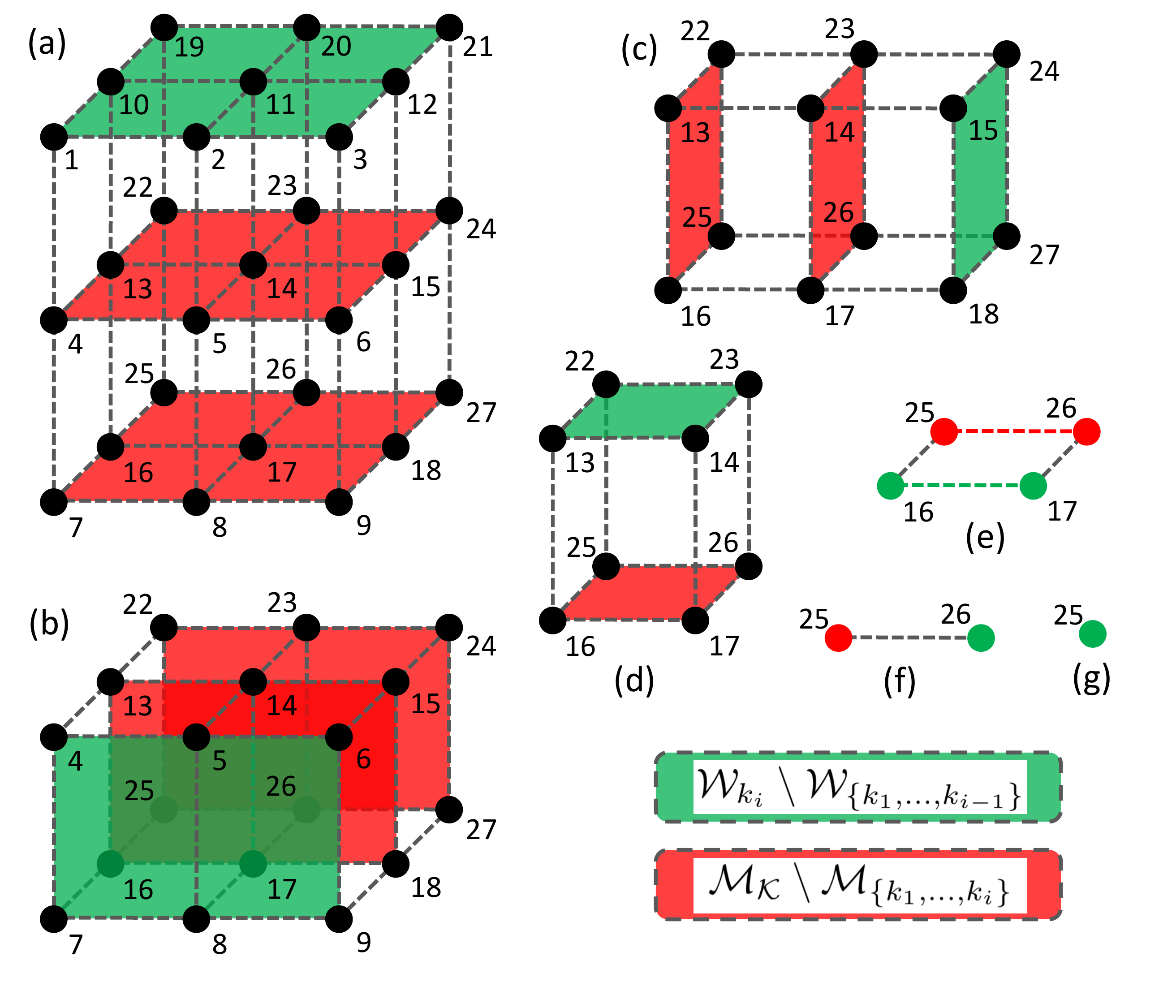}
\vspace{-0.5cm}
\caption{~\small A representation of  Example \ref{ex:optimality}
for a given permutation $\{1,7,6,2,8,5,3,9,4\}$. 
In (a)-(g), the $i$-th subfigure shows the functions of interests and  files of interests for node $k_i$, highlighted in green and red, respectively. For instance, (c) shows the functions and files of interests for node $6$, after accounting for functions and files of interests for node $1$ (see (a)) and node  $7$ (see (b)). 
From (a)-(g), $i$ increases by 1 at each step, and the lattice shrinks in one dimension by one unit. Refer to Fig. \ref{fig: 3d fig_sgt1} for the definition of file mapping and function assignment.}
\label{fig: 3d_proof_fig}
\vspace{-0.4cm}
\end{figure}
Similarly, for node $7$, we are count functions it computes and files it does not have locally available, except this time we do not count files available to node $1$ or functions assigned to node $1$. Fig.~\ref{fig: 3d_proof_fig}(b) shows the files and functions we are counting. Note that, we disregard the top layer of the cube which represents the files and functions assigned to node $1$. We see that
$H\left(V_{\mathcal{W}_{7},:}|V_{:,\mathcal{M}_{7}},Y_{\{1 \}}\right) = T\cdot 6 \cdot 12 = 72T.$
By continuing this process, from Fig.~\ref{fig: 3d_proof_fig}(c-f), we see that
$H\left(V_{\mathcal{W}_{6},:}|V_{:,\mathcal{M}_{6}},Y_{\{1,7 \}}\right) = 32T,
H\left(V_{\mathcal{W}_{2},:}|V_{:,\mathcal{M}_{2}},Y_{\{1,7,6 \}}\right) = 16T,  
H\left(V_{\mathcal{W}_{8},:}|V_{:,\mathcal{M}_{8}},Y_{\{1,7,6,2 \}}\right) = 4T,   
H\left(V_{\mathcal{W}_{5},:}|V_{:,\mathcal{M}_{5}},Y_{\{1,7,6,2,8 \}}\right) = T.$
Finally, 
only $1$ lattice point remains in Fig.~\ref{fig: 3d_proof_fig} (g),  representing a function assigned to node $3$. However,  there are no lattice points representing files node $3$ does not have locally available. This occurs because the other two nodes aligned along the same dimension, nodes $1$ and $2$, have already been accounted for, and they collectively have all the files that node $3$ does not have. Therefore,
$H\left(V_{\mathcal{W}_{3},:}|V_{:,\mathcal{M}_{3}},Y_{\{1,7,6,2,8,5 \}}\right) = 0.$
Similarly, for the last two nodes of the permutation, nodes $9$ and $4$, there are no remaining files that are not locally available to them. In fact, there are also no functions assigned to nodes $9$ and $4$ which have not already been accounted for. Therefore,
$H\left(V_{\mathcal{W}_{9},:}|V_{:,\mathcal{M}_{9}},Y_{\{1,7,6,2,8,5,3 \}}\right) =
H\left(V_{\mathcal{W}_{4},:}|V_{:,\mathcal{M}_{4}},Y_{\{1,7,6,2,8,5,3,9 \}}\right) = 0.$

By taking the sum of (\ref{eq: bound_eq1}), we directly compute the bound of (\ref{eq: bound_prf_4}) and find that
\begin{align}
L^* &\geq \frac{287T}{QNT} = \frac{287}{27\cdot 27} \approx 0.3937.
\end{align}
\end{example}

\section{Discussions}
\label{sec: Discussion}
 In this section, we  compare the performance the proposed scheme
with the state-of-the-art scheme of  \cite{li2018fundamental} in terms of communication load and  required number of files and functions.
While the proposed design applies to heterogeneous network, the design in  \cite{li2018fundamental} only applies to homogeneous networks. Hence, to facilitate fair comparisons, we compare with an equivalent homogeneous network of   \cite{li2018fundamental} with the same  $r,N,Q$,
for appropriate choices of $\eta_1$ and $\eta_2$. 
The scheme of \cite{li2018fundamental} requires $N_1 = {K \choose r}\eta_1$ input files, $Q_1 = {K \choose s}\eta_2$ reduce functions, and achieves  the communication load as a function of $K$, $r$ and $s$ as \be \label{eq: LiResult}
L_1(r,s) = \sum_{\gamma = \max \{ r+1, s\}}^{\min \{ r+s, K \}} \frac{\gamma {K \choose \gamma} {\gamma-2 \choose r-1}{r \choose \gamma - s}}{r {K \choose r}{K \choose s}}.
\ee

\begin{corollary}
\label{corollary:optimality}
Let $L_{\rm c}(r)$ be the resulting communication load from using the NG File Mapping, Cascaded Function Assignment and MR Shuffle Method, and $L_1(r,r)$  given by (\ref{eq: LiResult}) for an equivalent computation load $r$ and number of nodes $K$ and $r=s$.
\begin{itemize}
\item[(a)] When $r=s=2$, for both homogeneous and heterogeneous hypercuboid designs, we have $L_c(2)<L_1(2,2)$. 
\item[(b)] When $r=s\ge 6$ and  $K>r-1+4r^3$, there exists a heterogeneous hypercuboid design where $L_{\rm c}(r)<L_1(r,r)$.
\item[(c)] In the limiting regime, when $r = s = o(K)$,\footnote{We will use the following standard ``order'' notation: given two functions $f$ and $g$, we say that: 1)  $f(K) = O\left(g(K)\right)$ if there exists a constant $c$ and integer $N$ such that  $f(K)\leq cg(K)$ for $n>N$. 2) $f(K)=o\left(g(K)\right)$ if $\lim_{K \rightarrow \infty}\frac{f(K)}{g(K)} = 0$.
3) $f(K) = \Omega\left(g(K)\right)$ if $g(K) = O\left(f(K)\right)$. 4)
$f(K) = \omega\left(g(K)\right)$ if $g(K) = o\left(f(K)\right)$.
5) $f(K) = \Theta\left(g(K)\right)$ if $f(K) = O\left(g(K)\right)$ and~$g(K) = O\left(f(K)\right)$.} we have
$
\lim_{K \rightarrow \infty}\frac{L_{\rm c}(r)}{L_1(r,r)} \leq 1.
\label{eq:Thm5}
$
\end{itemize}
\end{corollary}
\begin{IEEEproof}
Corollary \ref{corollary:optimality} is proven in Appendix \ref{sec: pf_homo_compare}.
\end{IEEEproof}
\subsection{Homogeneous Cascaded CDC} 
\label{sec: disc_sg1}
In this section, we provide numerical results to confirm the findings in Corollary \ref{corollary:optimality} for homogeneous cascaded CDC.  

\begin{figure}
\centering
\begin{subfigure}{.5\textwidth}
  \centering
  \includegraphics[width=8cm]{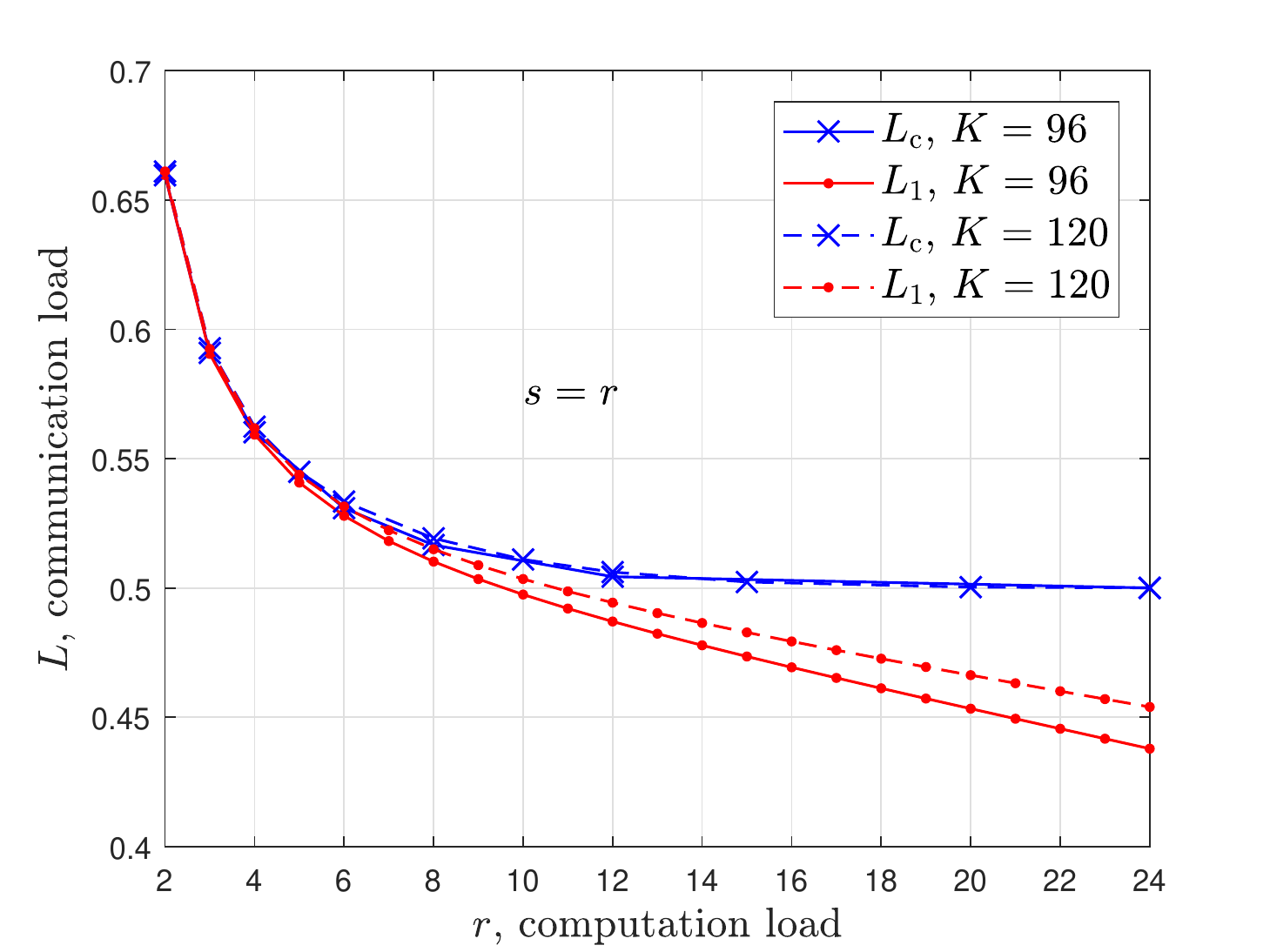}
  \captionof{figure}{Increase $r=s$ for fixed $K=96, 100$.}
\label{fig: srgraph}
\end{subfigure}%
\begin{subfigure}{.5\textwidth}
  \centering
  \includegraphics[width=8cm]{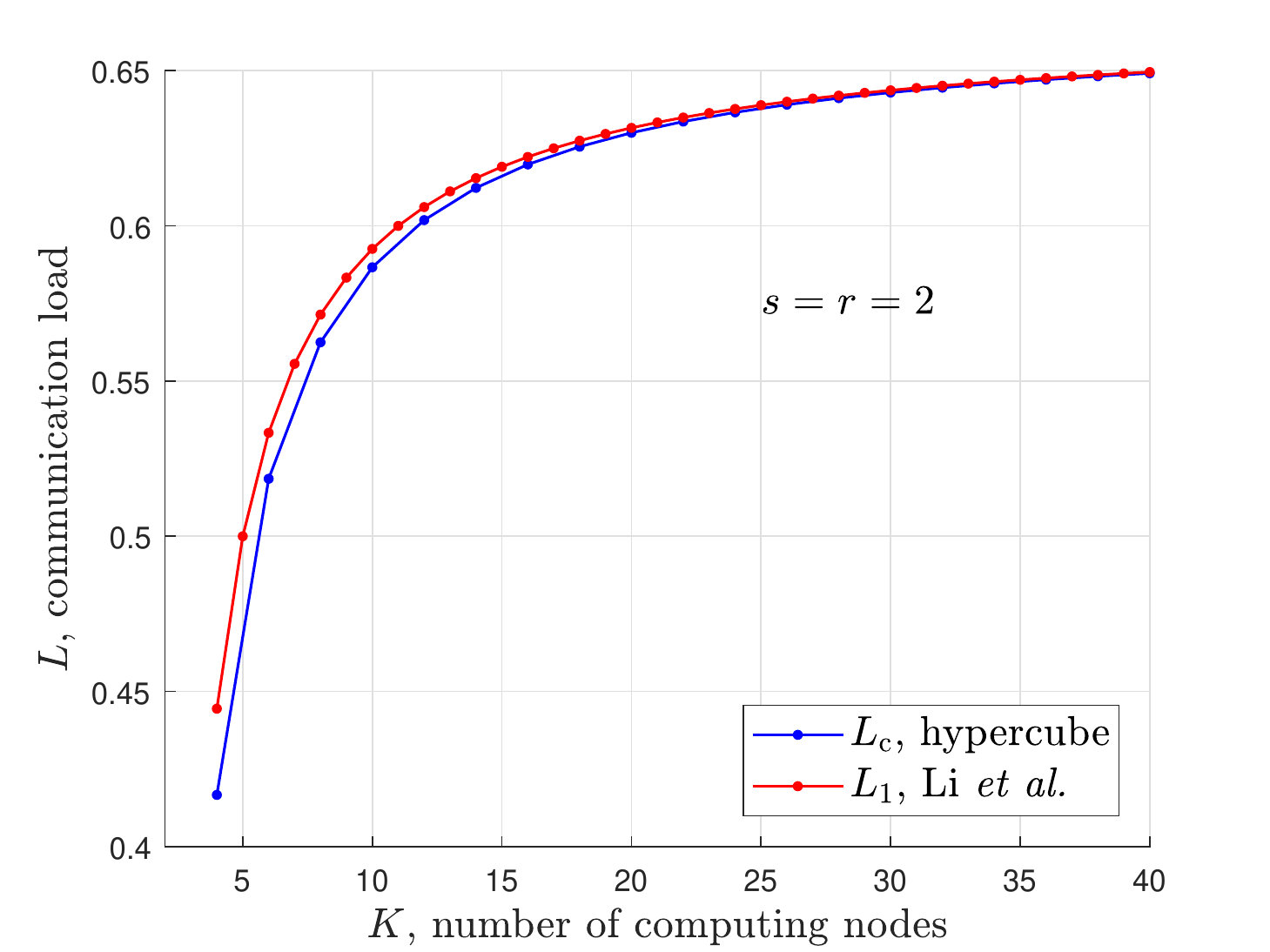}
 \captionof{figure}{Increase $K$ for fixed $r=s=2$.}
\label{fig: sr2graph}
\end{subfigure}
\vspace*{-0.3cm}
\caption{~\small Comparisons of  communication load $L_c$ of the proposed design and $L_1$ of \cite{li2018fundamental} for homogeneous networks.
}
\vspace{-0.75cm}
\end{figure}

In Fig. \ref{fig: srgraph}, we compare $L_c(r)$ with $L_c(r,r)$ for large homogeneous networks ($K=96, 120$) as $r$ increases. 
  For  $s=r$, we observe that $L_c(r)$ with $L_c(r,r)$ are close  
when $r<<K$, verifying
Corollary \ref{corollary:optimality} (c), but
\begin{comment}
\begin{theorem}
\label{theorem: homogeneous optimality}
Let $L_{\rm c}$ be the resulting communication load from usingthe NG File Mapping, Cascaded Function Assignment and MR Shuffle Method. 
When $r = s = o(K)$, then 
\be
\frac{L_{\rm c}}{L_1} = 1 + o(1).
\ee
where $L_1$ is given by (\ref{eq: LiResult}) for an equivalent computation load $r$.
\hfill$\square$
\end{theorem}
Theorem \ref{theorem: homogeneous optimality} is proven in Appendix \ref{sec: pf_homo_compare}.
\end{comment}
begin to deviate when ${r = \Theta (K)}$.
We see that for most (but not all) values of $m$ and $r$ that $L_{\rm c} = \frac{1}{2}+\epsilon$ where $\epsilon > 0$. The intuition behind this is for most of the Shuffle phase, IVs are included in coded pairs. Meanwhile, 
from (\ref{eq: LiResult}) and Fig. \ref{fig: srgraph}, 
we see that $L_1(r,r)$ can have a communication load less than $\frac{1}{2}$.
%

 Fig. \ref{fig: sr2graph} compares $L_c(r)$ and $L_1(r,r)$ as a function of $K$ for fixed $r=s=2$. This corresponds to  the limiting regime of $r=o(K)$. Moreover, consistent with Corollary \ref{corollary:optimality} (a), Fig. \ref{fig: sr2graph} shows the proposed design achieves a lower communication load than that  of \cite{li2018fundamental}. 
This is because while both the proposed scheme and that of \cite{li2018fundamental} handle IVs that are requested by $1$ or $2$ nodes with the same efficiency,  the former  has a greater fraction of IVs which are requested by $0$ nodes. 
The optimality of the scheme in  \cite{li2018fundamental} is proved under the key assumption on function assignment that 
 every $s$ nodes have at least 1 function in common. In contrast, we do not make such an assumption in the proposed design. This allows greater flexibility in the design of function assignment and enables a lower communication load than that of \cite{li2018fundamental}.
 

By the proposed NG File Mapping and Cascaded Function Assignment, the minimum requirement of $N$ and $Q$ is $\left( \frac{K}{r} \right)^{r}$ where $\eta_1 = \eta_2 = 1$. While the minimum requirements of $N$ and $Q$ in \cite{li2018fundamental} are ${K \choose r}$ and ${K \choose s}$. Hence, 
it can be observed that the proposed approach reduces the required numbers of both $N$ and $Q$ exponentially as a function of $r$ and $s$.

\subsection{Heterogeneous Cascaded CDC} 
We consider the following two cases of heterogeneous network. 
\begin{itemize}
  \item \textit{Case 1}: Assume $\frac{2}{5}$ of the nodes have $3$ times as much storage capacity and computing power compared to the other $\frac{3}{5}$ of the nodes. {Here,  we set $P=2, r_1=2, m_1=0.2 K, r_2=1, m_2=0.6K$}.  Note that $m_2=3 m_1$.
  \item \textit{Case 2}: Assume $\frac{1}{5}$ of the nodes have $4$ times as much storage capacity and computing power compared to the other $\frac{4}{5}$ of the nodes. Here, we set
   {  $P=2$, $r_1=2,  m_1=0.1 K, r_2=2, m_2=0.4K$ }. Note that $m_2=4m_1$.
\end{itemize}
We compare these two cases to  equivalent homogeneous schemes 
including the homogeneous scheme described in this paper and  the scheme of \cite{li2018fundamental}. Here, equivalent means the schemes are compared with the same $r$, $s$ and $K$.
\begin{figure}
\centering
\begin{subfigure}{.5\textwidth}
  \centering
  \includegraphics[width=8cm]{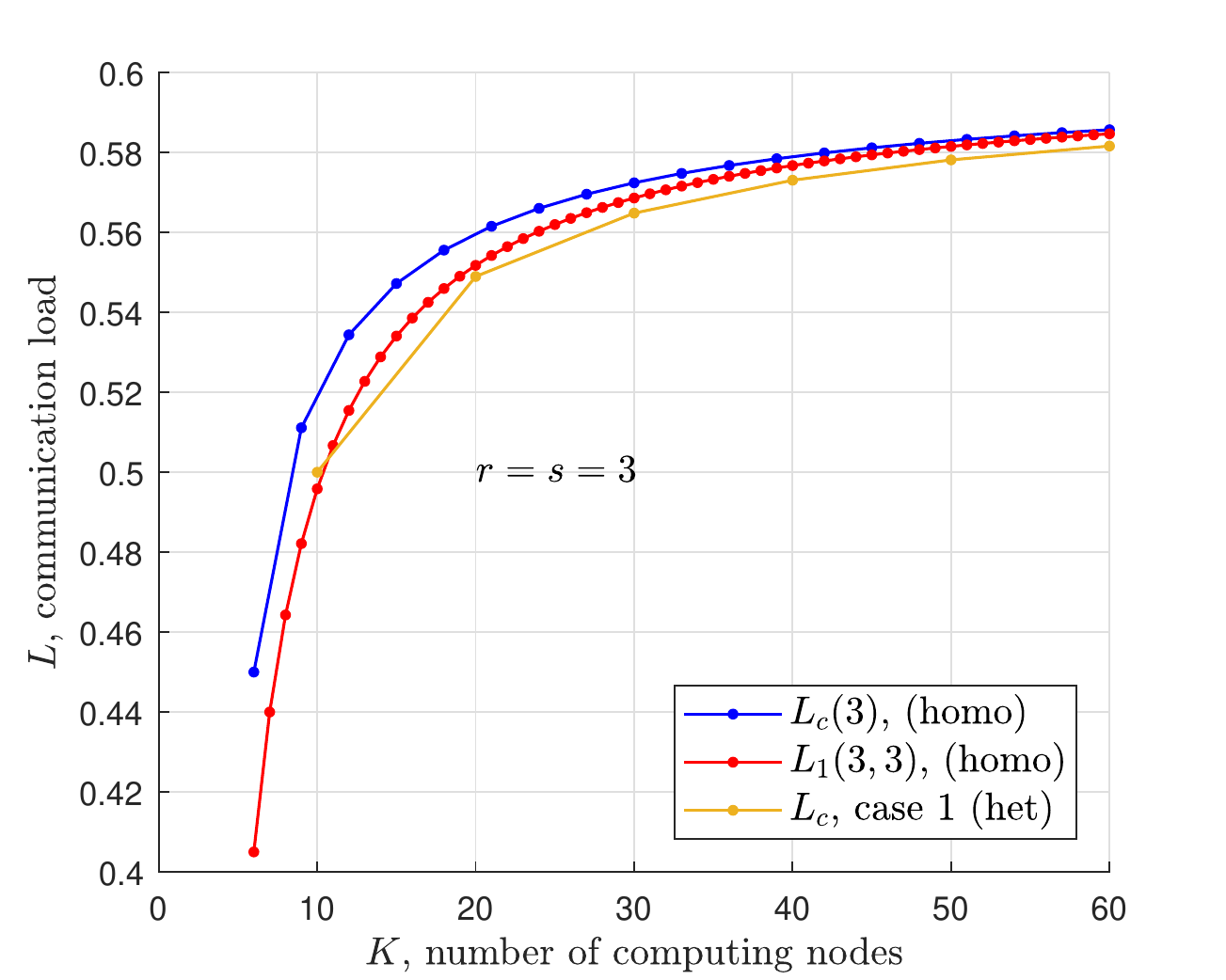}
 \captionof{figure}{Case 1}
\end{subfigure}%
\begin{subfigure}{.5\textwidth}
  \centering
  \includegraphics[width=8cm]{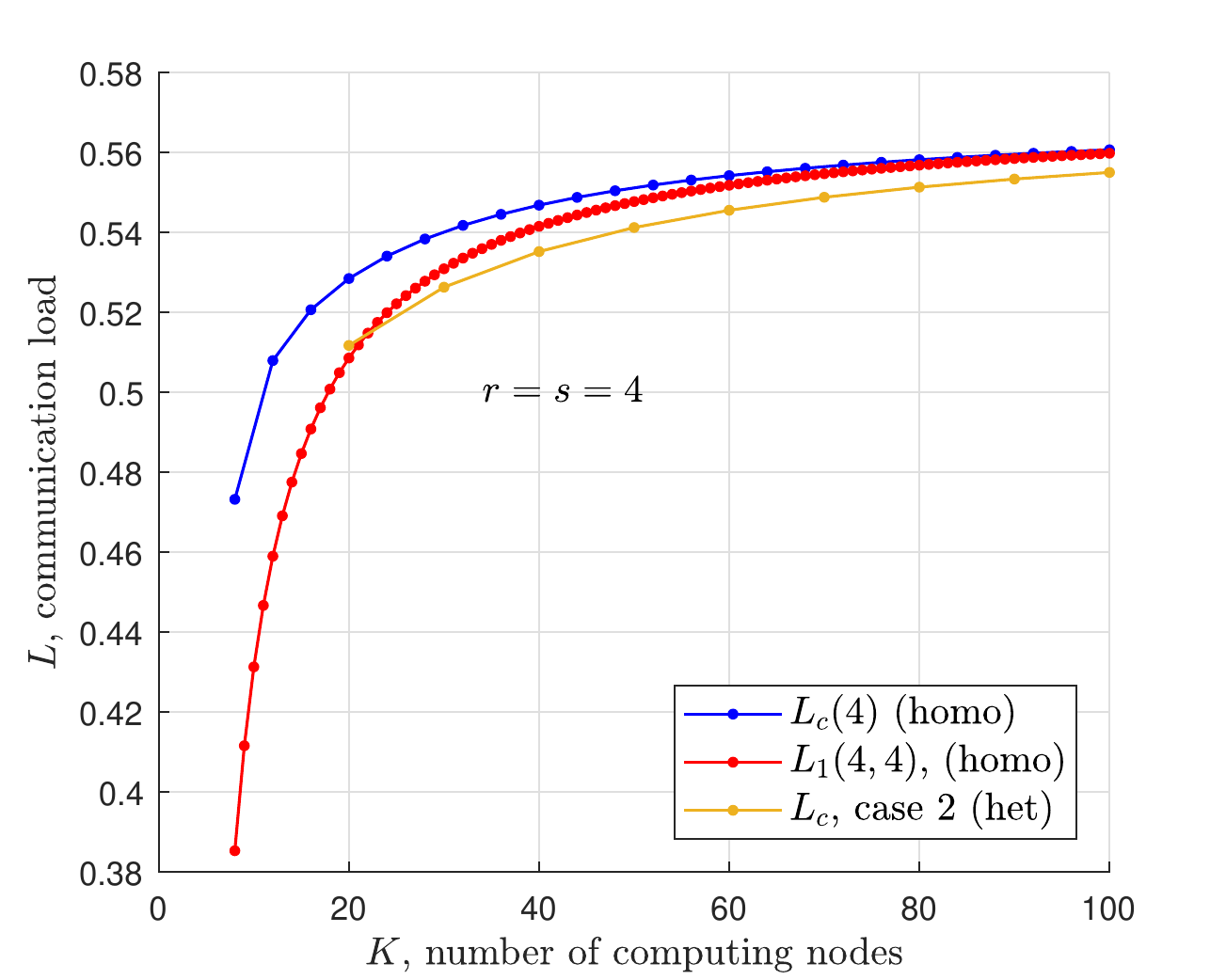}
 \captionof{figure}{Case 2}
\end{subfigure}
\vspace*{-0.5cm}
\caption{~\small Comparisons of the communication load achieved by the proposed heterogeneous design to equivalent homogeneous designs including the proposed design and the design from \cite{li2018fundamental}.}
\vspace{-0.3cm}
\label{fig: het sim_fig}
\end{figure}
 Fig.~\ref{fig: het sim_fig} confirms {Corollary \ref{corollary:optimality} (b) that for fixed $r$ and large  $K$, there exists a proposed heterogeneous design with $L_c(r)<L_c(r,r)$}.
There appears to be an advantage of having a set of nodes with both more locally available files and assigned functions. In this way, less IV shuffling is required to satisfy the requests of these nodes. As discussed before, an extreme case of this can be observed where a subset of nodes each have all files locally available and compute all assigned functions. Furthermore, for the given simulations, the communication load of the heterogeneous designs approaches the communication load of the homogeneous designs {as shown in Corollary \ref{corollary:optimality} (c)}. 
\begin{comment}
\begin{theorem}
\label{theorem: heterogeneous ub}
For heterogeneous coded cascaded distributed computing, 
when $r=o(K)$, 
then {\rr In (21) do we really need $r \rightarrow \infty?$ Should it be $K \rightarrow \infty$ and $r=o(K)$? This should include the case when $r$ is fixed, just like Fig. 6. }
\be
\lim_{r \rightarrow \infty}\frac{L_{\rm c}(r)}{L_1(r,r)} \leq 1 + o(1).
\label{eq:Thm5}
\ee
\end{theorem}
\begin{IEEEproof}
From (\ref{eq:L_c_upper}),
we have $L_{\rm c}  \le \frac{r}{2r-1}$. Moreover, it was shown in (\ref{eq: L1_low_bound}) that $L_1 > \frac{r}{2r-1}\cdot (1 + o(1))$ when $r=o(K)$. Thus, (\ref{eq:Thm5}) holds.
\end{IEEEproof}
\end{comment}

\section{Conclusion}
\label{sec: Conclusion}

In this work, we introduced a novel combinatorial hypercuboid approach for cascaded CDC frameworks with both homogeneous and heterogeneous network scenarios. The proposed low complexity combinatorial structure can determine both input file and output function assignments, requires significantly less number of input files and output functions, and operates on large heterogeneous networks where nodes have varying storage space and computing resources. Surprisingly, due to a different output function assignment, the proposed scheme can outperform the optimal state-of-the-art scheme with a different output function assignment. Moreover, we also show that the heterogeneous storage and computing resource can reduce the communication load compared to its homogeneous counterpart.
Finally, the proposed scheme can be shown to be optimal within a constant factor of the information theoretic converse bound while fixing the input file and the output function assignments.


\appendices


\if{0}
\section{Proof of Theorem \ref{theorem: bound}}
\label{appendix: bound}
In this proof, we use the following notation: $\mathcal{K}$ is the set of all nodes, $X_{\mathcal{K}}$ represents the collection of all transmissions by all nodes in $\mathcal{K}$, $\mathcal{W}_{\mathcal{S}}$ is the set of functions assigned to at least on node of $\mathcal{S}$, $\mathcal{M}_{\mathcal{S}}$ is the set files locally available to at least one node in $\mathcal{S}$, $V_{\mathcal{W}_{\mathcal{S}_1},\mathcal{M}_{\mathcal{S}_2}}$ is the set of intermediate values needed to compute the functions of $\mathcal{W}_{\mathcal{S}_1}$ and computed from the files of $\mathcal{M}_{\mathcal{S}_2}$. Finally, we define the following
\be
Y_\mathcal{S} \triangleq \left(V_{\mathcal{W}_\mathcal{S},:},V_{:,\mathcal{M}_\mathcal{S}}\right)
\ee
where ``$:$" is used to denote all possible indices.

%
Given all the transmissions from all nodes, $X_\mathcal{K}$, and intermediate values which can be locally computed by a node $k$, $V_{:,\mathcal{M}_k}$, node $k$ needs to have access to all intermediate values necessary for its assigned functions, $V_{\mathcal{W}_k,:}$, therefore
\be
H(V_{\mathcal{W}_k,:} | X_\mathcal{K}, V_{:,\mathcal{M}_k}) = 0.
\ee

Given this assumption, we find
\begin{align}
H(X_\mathcal{K}) &\geq H(X_\mathcal{K}|V_{:,M_{k_1}}) \nonumber\\
&= H(X_\mathcal{K},V_{\mathcal{W}_{k_1},:}|V_{:,M_{k_1}}) - H(V_{\mathcal{W}_{k_1},:} | X_\mathcal{K}, V_{:,\mathcal{M}_{k_1}}) \nonumber\\
&= H(X_\mathcal{K},V_{\mathcal{W}_{k_1},:}|V_{:,M_{k_1}}) \nonumber\\
&= H(V_{\mathcal{W}_{k_1},:}|V_{:,M_{k_1}}) + H(X_\mathcal{K}|V_{\mathcal{W}_{k_1},:},V_{:,M_{k_1}})\nonumber\\
&=H(V_{\mathcal{W}_{k_1},:}|V_{:,M_{k_1}}) + H(X_\mathcal{K}|Y_{k_1}). \label{eq: claim 1 2}
\end{align}
Similarly,
\begin{align}
H&(X_\mathcal{K}|Y_{\{k_1,\ldots k_{i-1}\}}) \nonumber\\
&\geq H(X_\mathcal{K}|V_{:,M_{k_i}},Y_{\{k_1,\ldots k_{i-1}\}}) \nonumber\\
&= H(X_\mathcal{K},V_{\mathcal{W}_{k_i},:}|V_{:,M_{k_i}},Y_{\{k_1,\ldots k_{i-1}\}}) \nonumber\\
&\text{  }\text{ }\text{ }- H(V_{\mathcal{W}_{k_i},:} | X_\mathcal{K}, V_{:,\mathcal{M}_{k_i}},Y_{\{k_1,\ldots k_{i-1}\}}) \nonumber\\
&= H(X_\mathcal{K},V_{\mathcal{W}_{k_i},:}|V_{:,M_{k_i}},Y_{\{k_1,\ldots k_{i-1}\}}) \nonumber\\
&= H(V_{\mathcal{W}_{k_i},:}|V_{:,M_{k_i}},Y_{\{k_1,\ldots k_{i-1}\}}) \nonumber \\
&\text{ }\text{ }\text{ }+ H(X_\mathcal{K}|V_{\mathcal{W}_{k_i},:},V_{:,M_{k_i}},Y_{\{k_1,\ldots k_{i-1}\}})\nonumber\\
&= H(V_{\mathcal{W}_{k_i},:}|V_{:,M_{k_i}},Y_{k_1,\ldots k_{i-1}}) + H(X_\mathcal{K}|Y_{\{k_1,\ldots k_i\}}). \label{eq: claim 1 3}
\end{align}
Also, since nodes can only transmit intermediate values from locally available files, we see that
\be
H(X_\mathcal{K}|Y_{\{k_1,\ldots k_K\}}) = 0.
\ee
By starting with (\ref{eq: claim 1 2}) and iteratively using the relationship of (\ref{eq: claim 1 3}) to account for all $k_i \in \mathcal{K}$, we obtain
\be \label{eq: th 4 ent sum}
H(X_\mathcal{K}) \geq \sum_{i=1}^{K} H\left(V_{\mathcal{W}_{k_i},:}|V_{:,\mathcal{M}_{k_i}},Y_{\{k_1,\ldots, k_{i-1} \}}\right).
\ee
Moreover, since $H(X_\mathcal{K}) = LTQN$, from (\ref{eq: th 4 ent sum}) we obtain the lower bound on the optimal communication load, $L^*$, of (\ref{eq: bound_eq1}) and prove Theorem \ref{theorem: bound}.
\fi

\section{Proof of Theorem  \ref{theorem: sg1het}}
\label{sec: achieve sg1}
  { Let $x_i = |\mathcal{K}_i|$ be the size of the $i$-th dimension of the hypercuboid. Note that if $\mathcal K_i \in \mathcal C_p$, then $x_i=m_p$, as defined in General Node Grouping of Section \ref{sec: gen_sd}.}
The communication load can be calculated by considering all $r$ rounds of the Shuffle phase. For $\gamma\in\{1,\ldots,r-1\}$, in the $\gamma$-th round we use the IG Shuffle Method. We consider a node group $\mathcal{S}$ of $2\gamma$ nodes where there are $2$ nodes from $\mathcal{K}_i$ for all $i \in \mathcal{A}\subseteq [r]$ such that $|\mathcal{A}|=\gamma$.  Given $\mathcal{A}$ and $\mathcal{S}$ we identify all node sets $\mathcal{Y}$ which contain $r-\gamma$ nodes, $1$ node from each set $\mathcal{K}_i$ for all $i \in [r]\setminus \mathcal{A}$. Given $\mathcal{A}$, there are
$\prod\limits_{i\in [r]\setminus \mathcal{A}}x_i$
 possibilities for $\mathcal{Y}$.
Furthermore, there are $2^\gamma$ possibilities for choosing a subset $\mathcal{S}'\subset \mathcal{S}$ such that $|\mathcal{S}'|=\gamma$.
Therefore, there are
\begin{align}
2^\gamma\prod_{i\in\mathcal{A}}{x_i \choose 2}\prod_{i \notin \mathcal{A}}x_i &= 2^\gamma\prod_{i\in\mathcal{A}}\frac{x_i(x_i-1)}{2}\prod_{i \notin \mathcal{A}}x_i = X \prod_{i\in\mathcal{A}} (x_i-1)
\end{align}
unique pairs of $\mathcal{Y}$ and $\mathcal{S}'$ given $\mathcal{A}$. For each unique pair of $\mathcal{Y}$ and $\mathcal{S}'$, we define a set of IVs $\mathcal{V}_{\mathcal{S}'\cup \mathcal{Y}}^{\mathcal{S}\setminus \mathcal{S}'}$ which only contains IVs $v_{i,j}$ such that $i\in\mathcal{D}_\alpha$ and $w_j\in\mathcal{B}_\ell$ where $\{\{ \mathcal{S}\setminus\mathcal{S}'\}\cup \mathcal{Y} \}=\mathcal{T}_\alpha$ and $\mathcal{S}' \cup \mathcal{Y} = \mathcal{T}_\ell$. Since $|\mathcal{B}_\ell|=\eta_1$ and $|\mathcal{D}_\alpha|=\eta_2$, we see that $|\mathcal{V}_{\mathcal{S}'\cup \mathcal{Y}}^{\mathcal{S}\setminus \mathcal{S}'}|=\eta_1 \eta_2$. All of the IV sets are transmitted in coded pairs, effectively reducing the contribution to the communication load by half. Therefore, given $\mathcal{A}$, there are
$\frac{\eta_1\eta_2X}{2} \prod_{i\in\mathcal{A}} (x_i-1)$
transmissions of size $T$ bits, the number of bits in a single IV. $\mathcal{A}$ can range in size from $1$ to $\gamma-1$. Accounting for all possibilities of $\mathcal{A}$ and normalizing by $QNT=\eta_1\eta_2X^2T$, we obtain the number of bits transmitted as
\be
\label{eq: shuffmeth2}
\frac{\eta_1\eta_2T}{2X}\sum_{\gamma = 1}^{r-1}\left(\sum_{\substack{ \{\mathcal{A}:\mathcal{A}\subset \left[r\right],  |\mathcal{A}|=\gamma\}}}\left(\prod_{i\in\mathcal{A}}\left( x_i-1\right)\right)\right).
\ee

Finally, in the $r$-th round, we use the LC Shuffle Method. We consider all node groups of $2r$ nodes, $\mathcal{S}$, such that $|\mathcal{S}\cap \mathcal{K}_i|=2$ for all $i\in [r]$. There are
$\prod_{i=1}^{r}{x_i \choose 2}$
possibilities for a node group $\mathcal{S}$. Furthermore, given $\mathcal{S}$, there are $2^r$ possibilities for a node group $S'\subset\mathcal{S}$ such that $|\mathcal{S}'\cap \mathcal{K}_i|=1$ for all $i\in [r]$. We see that $\mathcal{S}'=\mathcal{T}_\ell$ and $\{ \mathcal{S}\setminus\mathcal{S}' \}=\mathcal{T}_{\alpha_\ell}$ for some $\ell$ which determines $\alpha_\ell$. Therefore, $|\mathcal{V}_{\mathcal{S}'}^{\mathcal{S}\setminus \mathcal{S}'}|=|\mathcal{B}_\ell|\cdot|\mathcal{D}_{\alpha_\ell}|=\eta_1 \eta_2$. Each node of $\mathcal{S}$ transmits $2^{r-1}$ linear combinations of size
$\frac{\eta_1\eta_2T}{2r-1}$
bits and the total number of bits transmitted in the $r$-th round is
\be
\label{eq: shuffmeth3}
\frac{2r\eta_1\eta_2 2^{r-1}T}{2r-1}\prod_{i=1}^{r}{x_i \choose 2} = \frac{r\eta_1\eta_2TX}{2r-1}\prod_{i=1}^{r}(x_i -1).
\ee
Next, we need to add (\ref{eq: shuffmeth2}),  (\ref{eq: shuffmeth3}), and normalize by $QNT=\eta_1\eta_2X^2T$ to get $L_c$. 
The summation can be simplified using Lemma \ref{lm: sum_of_products} below. 
\begin{lemma}\label{lm: sum_of_products}
  Given a set of numbers $a_1,a_2,\ldots,a_c\in \mathbb{R}$, the sum of the product of all subsets, including the empty set, of this set of numbers is
  \be
  \label{eq:lemma2_sum}
  \sum_{\mathcal{C}\subseteq [c]} \prod_{i\in \mathcal{C}} a_i = (a_1 + 1)\times(a_2+1)\times \cdots \times (a_c +1).
  \ee
\end{lemma}
Lemma \ref{lm: sum_of_products} easily follows
by considering the expansion of the right side of  
(\ref{eq:lemma2_sum}).  
Using Lemma \ref{lm: sum_of_products}, the communication load (\ref{eq: eqL_sgt1}) is given by 
\begin{align}\label{eq: eqiv_eq_L}
L_{\rm c} &= \frac{1}{2X}\left( \prod_{i=1}^{r} |\mathcal{K}_i| - 1 - \prod_{i=1}^{r} (|\mathcal{K}_i|-1) \right) + \frac{r}{X(2r-1)}\prod_{i=1}^{r} (|\mathcal{K}_i|-1) \nonumber \\
&= \frac{X-1}{2X} + \frac{1}{X(4r-2)}\prod_{i=1}^{r} (|\mathcal{K}_i|-1) < \frac{1}{2} +  \frac{1}{4r-2}.
\end{align}

\section{Correctness of Heterogeneous 
CDC Scheme }
\label{app:correctness}
Consider $r=s$ sets of IVs, where  the $\gamma$-th set includes IVs  requested by $\gamma$ nodes. For each set, we prove that Shuffle Methods from Section \ref{sec: gen_sd} satisfy the following: 1) all IVs from that set are included in a coded transmission, 2) nodes can decode IVs they request from that set and 3) nodes only transmit IVs from that set which are computed from locally available files. Then by using the specified Shuffle Method for each $\gamma \in \{1,\ldots ,s \}$, each node will receive all its requested IVs and be able to compute all assigned functions in the Reduce phase.

We first prove criterion 1) for the IG and LC Shuffle Methods. For $\gamma\in\{1,\ldots,r-1 \}$, in the $\gamma$-th round we see $|\mathcal{T}_\alpha \cap \mathcal{T}_\ell|=|\mathcal{Y}|=r-\gamma.
$
 Also, any $\mathcal{T}_\ell$ is possible and given $\mathcal{T}_\ell$ any $\mathcal{T}_\alpha$ is possible given that $|\mathcal{T}_\alpha \cap \mathcal{T}_\ell|=r-\gamma$. Therefore, the set of IVs transmitted is
\be
 \{v_{i,j} : i \in \mathcal{D}_\alpha, w_j \in \mathcal{B}_\ell, |\mathcal{T}_\ell \cap \mathcal{T}_\alpha| = r-\gamma \}.
\ee
This is the set of all IVs requested by $\gamma$ nodes and this proves 1) for the IG Shuffle Method. Similarly, for the $r$-th round, in the LC Shuffle Method, we consider all possible pairs $\mathcal{T}_\ell$ and $\mathcal{T_{\alpha\ell}}$ such that $|\mathcal{T}_\alpha \cap \mathcal{T}_\ell|=0$ and the sets have no nodes in common. The IVs included in the linear combinations in the $r$-th are then
$
 \{v_{i,j} : i \in \mathcal{D}_{\alpha_\ell}, w_j \in \mathcal{B}_\ell, |\mathcal{T}_\ell \cap \mathcal{T}_{\alpha_\ell}| = 0 \}.
$
which represents all IVs requested by $r$ nodes and this proves 1) for the LC Shuffle Method.

Next, for the IG Shuffle Method, consider an arbitrary node $z \in \mathcal{S}$ that receives a multicast message from node $y \in \mathcal Y$ where $z\notin\mathcal{Y}$. The message is of the form
$\mathcal{V}_{\mathcal T_\ell}^{\mathcal{S}\setminus \mathcal{S}'} \oplus \mathcal{V}_{\mathcal T_\alpha}^{\mathcal{S}'}$, given in (\ref{eq:def_message_shffule2}), where
$\mathcal{T}_\alpha = \{\mathcal{S}\setminus \mathcal{S}'\}\cup \mathcal{Y}$ and  $\mathcal{T}_\ell =  \mathcal{S}'\cup\mathcal{Y}$.  Note that $z$ is either in $\mathcal{S}'$ or $\mathcal{S}\setminus\mathcal{S}'$.  If  $z \in \mathcal{S}'$, then since $z \in \mathcal T_\ell$, it  has access to $\mathcal B_\ell$ and thus can compute all IVs in  $\mathcal{V}_{\mathcal T_\ell}^{\mathcal{S} \setminus \mathcal{S}'}$ and then subtract these off from the coded message to recover its desired IVs in  $\mathcal{V}_{\mathcal T_\alpha}^{\mathcal{S}'}$. The same reasoning applies to the case when $z \in \mathcal{S}\setminus\mathcal{S}'$.  This confirms 2). To confirm 3), we see that for any node $y \in \mathcal Y$, since $y$ is in both $\mathcal T_\ell$  and $\mathcal T_\alpha$, by the NG File Mapping node $y$ has access to both $\mathcal B_\ell$  and $\mathcal B_\alpha$ and thus can compute IVs in both $\mathcal{V}_{\mathcal T_\ell}^{\mathcal{S}\setminus \mathcal{S}'} $ and $ \mathcal{V}_{\mathcal T_\alpha}^{\mathcal{S}'}$.

For the LC Shuffle Method in the $r$-th round, for a given $\mathcal{S}$, there are $2^r$ choices of
$\mathcal T_\ell$ in (\ref{eq:ivs_shuffle3}) which determines the node group $\mathcal T_{\alpha_\ell}$. Fix a node $z \in \mathcal S$. Since half of these $\mathcal T_\ell$ include node $z$, we see that $z$ can compute exactly half of these IVs, and requests the other half of them. These leads to $2^{r-1}$ unknown IV sets $\mathcal{V}_{\mathcal T_{\ell}}^{\mathcal T_{\alpha_{\ell}}}$ requested by node $z$. Since
These IVs are further divided into $2r -1$ disjoint subsets, node $z$ will request a total of $(2r-1)2^{r-1}$ unknown packets. Node $z$ will receive transmissions from the other $2r - 1$ nodes in $\mathcal{S}$ in which each node transmits $2^{r - 1}$ linear combinations of its known IV sets of interest. Therefore, node $z$ can recover the $(2r-1)2^{r-1}$ unknown packets since it receives $(2r-1)2^{r-1}$ linear combinations. 
This proves criterion 2) for the LC Shuffle Method. To confirm 3) for LC Shuffle Method, we see that since node $k \in \mathcal T_\ell$, it has access to $\mathcal B_\ell$, and thus can compute all IVs in $\mathcal{V}_{\mathcal T_{\ell}}^{\mathcal T_{\alpha_{\ell}}}$.

\section{Proof of Theorem \ref{theorem: sg1_homo_bound}}
\label{sec: lower bound}

The proof of Theorem \ref{theorem: sg1_homo_bound} utilizes Lemma \ref{theorem: bound} in 
  \cite{woolsey2020cdc} 
which is based on the approaches in \cite{wan2020index,li2018fundamental} and provides a lower bound on the entropy of all transmissions in the Shuffle phase given a specific function and file placement and a permutation of the computing nodes. 
\begin{lemma}
\label{theorem: bound}
Given a particular file placement  and function assignment \{$\mathcal{M}_k, \mathcal{W}_k, \;\forall k\in[K]\}$, in order for every node $k\in[K]$ to have access to all IVs necessary to compute functions of $\mathcal{W}_k$, the optimal communication load over all achievable shuffle schemes, $L^*$, is bounded by
  \be \label{eq: bound_eq1}
  L^* \geq  \frac{1}{TQN}\sum_{i=1}^{K} H\left(V_{\mathcal{W}_{k_i},:}|V_{:,\mathcal{M}_{k_i}},Y_{\{k_1,\ldots, k_{i-1} \}}\right)
  \ee
  where $k_1, \ldots , k_K$ is some permutation of $[K]$, $V_{\mathcal{W}_{k_i},:}$ 
  is the set of IVs necessary to compute the functions of $\mathcal{W}_{k_i}$. Here, {the notation ``$:$" is used to denote all possible indices.} $V_{:,\mathcal{M}_{k_i}}$ is set of IVs which can be computed from the file set $\mathcal{M}_{k_i}$ and $Y_{\{k_1,\ldots, k_{i-1} \}}$  is the union of the set of IVs necessary to compute the functions of  $\bigcup_{j=1}^{i-1}\mathcal{W}_{k_j}$ and the set of IVs which can be computed from files of $\bigcup_{j=1}^{i-1}\mathcal{M}_{k_j}$.
\hfill $\square$
\end{lemma}

{\bf Proof of Theorem \ref{theorem: sg1_homo_bound}}
\begin{comment}
{\nw To prove (\ref{eq:bound_homo_L*}), we use a generalized approach of Example 4. 
We pick a permutation of nodes } by first dividing the $K$ nodes into $m=\frac{K}{s}=\frac{K}{r}$ disjoint $([r],1)$ node groups $\mathcal{G}_1,\ldots, \mathcal{G}_m$, each containing a node from  $\{\mathcal{K}_i, i\in[r]\}$.
Then the  permutation is defined as
$\{ k_{\color{red}(j-1)s+1},\ldots, k_{js} \} = \mathcal{G}_j$
for all $j\in\{1,\ldots ,m \}$.
Given this permutation, we will show that
\begin{align}
\label{eq: bound_prf_1}
H&\left(V_{\mathcal{W}_{k_i},:}|V_{:,\mathcal{M}_{k_i}},Y_{\{k_1,\ldots, k_{i-1} \}}\right) = \eta_1\eta_2T\hat{m}^{\left(2s-2\ell\right)}(\hat{m}-1)^{(2\ell-1)}
\end{align}
where $\hat{m} = m - \lfloor \frac{i-1}{s}\rfloor$ and $\ell = i - s\lfloor \frac{i-1}{s} \rfloor$. By this labeling, $k_i$ is the $\ell$-th node of node set $\mathcal{G}_j$ where $j = m-\hat{m} + 1$.
\end{comment}
We pick a permutation of nodes by first dividing the $K$ nodes into $m=\frac{K}{r}$ disjoint $([r],1)$ node groups $\{\mathcal{G}_1,\ldots, \mathcal{G}_m\}$, each containing a node from  $\{\mathcal{K}_i, i\in[r]\}$.
Note that each $\mathcal K_i$ contains $m$ nodes aligned along $i$-th dimension of the hypercube, and each $\mathcal G_i$ has size $r$. In particular, each $\mathcal G_i, i\in [m]$ is an element in the set of all possible $([r],1)$ node groups $\{\mathcal T_1, \cdots, \mathcal T_X\}$ as defined in Single Node Grouping of Section \ref{sec: gen_sd} with $P=1$.
Then the permutation is defined such that $\mathcal{G}_1$ contains the first $r$ nodes, $\mathcal{G}_2$ contains the next $r$ nodes and this pattern continues such that $\mathcal{G}_m$ contains the last $r$ nodes of the permutation. In other words,
$\{ k_{(j-1)r+1},\ldots, k_{jr} \} = \mathcal{G}_j$
for all $j\in\{1,\ldots ,m \}$.
Given this permutation, to compute the $i$-th term of (\ref{eq: bound_eq1}), we will show
\begin{align}
\label{eq: bound_prf_1}
H&\left(V_{\mathcal{W}_{k_i},:}|V_{:,\mathcal{M}_{k_i}},Y_{\{k_1,\ldots, k_{i-1} \}}\right) = \eta_1\eta_2T\hat{m}_i^{\left(2r-2\ell_i\right)}(\hat{m}_i-1)^{(2\ell_i-1)}
\end{align}
where $\hat{m}_i = m - \lfloor \frac{i-1}{r}\rfloor$ and $\ell_i = i - r\lfloor \frac{i-1}{r} \rfloor$.  Note that nodes $\{k_1, k_2, \cdots, k_{i-1}\}$ consists of all nodes in $\{\mathcal G_1, \mathcal G_2, \cdots, \mathcal G_{\lfloor  \frac{i-1}{r}\rfloor}\}$, $\ell_i-1$ nodes in $ \mathcal G_{\lfloor  \frac{i-1}{r} \rfloor+1}$, and no nodes in any of the $\hat m_i -1$ node groups in $\{ \mathcal G_{\lfloor  \frac{i-1}{r}\rfloor +2}, \cdots, \mathcal G_m\}$.
In particular, $k_i$ is the $\ell_i$-th node in $\mathcal{G}_{n_i}$ where $n_i =\lfloor  \frac{i-1}{r}\rfloor + 1$. 

Since the IVs are assumed to be independent, we will take two steps to count the number of terms in (\ref{eq: bound_prf_1}). 
  In Step 1, we count the number of functions that are in $\mathcal W_{k_i}$, but not in $\{ \mathcal W_{k_1}, \cdots, \mathcal W_{k_{i-1}}\}$. These are referred to as functions of interests. 
  By the definition of cascaded function assignment, this is equivalent to counting the number of $([r],1)$ node groups $\mathcal T_l$ such that $\mathcal T_l$ includes $k_i$, but none of  nodes in $\{k_1, k_2, \cdots, k_{i-1}\}$. 
  Now, consider the {first} $\ell_i$ nodes in $\mathcal G_{n_i}$. Without loss of generality, assume that these nodes are taken from $\mathcal K_j, j=1, \cdots, \ell_i$, respectively. Then, for any dimension $r_0\in\{ \ell_i+1,\cdots,r \}$, $\mathcal T_{l,r_0}$ can be any of the $\hat m_i$ elements from $\{\mathcal G_{j,r_0}, n_i\le j \le m\}$. Here, $\mathcal T_{l,r_0}$ (or $\mathcal G_{j,r_0}$) denotes the element in $\mathcal T_{l}$  ( or $\mathcal  G_j$)  that is chosen from $\mathcal K_{r_0}$. Similarly, for any  $r_0\in\{ 1,\cdots,\ell_i-1 \}$, $\mathcal T_{l,r_0}$ can be any of the $\hat m_i-1$ elements from $\{\mathcal G_{j,r_0}, n_i+1\le j \le m\}$. When $r_0=\ell_i$, we must have $\mathcal T_{l,r_0}=k_i$.
  This gives a total of $ \hat{m}_i^{r-\ell_i}(\hat{m}_i-1)^{\ell_i-1}$ choices of such $\mathcal T_l$.
  In Step 2, we count the number of files that are  not in $\{ \mathcal M_{k_1}, \cdots, \mathcal M_{k_{i}}\}$. These are referred to as files of interests. This  step is equivalent to counting the number of  $([r],1)$ node groups  $\mathcal T_l$  that do not include any of the nodes $\{k_1, k_2, \cdots, k_{i}\}$. By replacing the case of $r_0 \in \{1,2, \cdots \ell_i-1\}$ in Step 1 by $r_0 \in \{1,2, \cdots \ell_i\}$, we obtain a total of  $\hat{m}_i^{r-\ell_i}(\hat{m}_i-1)^{\ell_i}$  choices of  $\mathcal T_l$. By taking the product of the results as in (\ref{eq:count})
  from both steps and accounting for the number of files, $\eta_1$, and functions, $\eta_2$, assigned to a node group $\mathcal{T}_l$, we obtain (\ref{eq: bound_prf_1}). 
  { The counting principle described above  can be visualized in Example 4. For instance, in Step 2,  when considering node $k_i$ after some ``layers have been peeled off'' (previous nodes were considered), the hypercuboid has $\ell_i$ dimensions of size $\hat{m}_i-1$ and $r-\ell_i$ dimensions of size $\hat{m}_i$.}

{ It  follows from  (\ref{eq: bound_eq1}) that we can  sum (\ref{eq: bound_prf_1}) over all nodes $\{k_i, i=1,\cdots, mr\}$ to calculate the lower bound corresponding to this permutation. 
Note that summing over the right side of (\ref{eq: bound_prf_1}) from $i=1$ to $i=mr$ is the same as summing over all possible $mr$ pairs of $(\hat m_i, \ell_i)$, where $\hat m_i$ goes from $1$ to $m$, and $\ell_i$ goes from $1$ to $r$. 
For instance, nodes in $\mathcal G_j$ all have the same $\hat m_i=m-j+1$ but different $\ell_i$ that goes from $1$ to $r$.  In the following, for brevity, we drop the subscript $i$ in the double summation over $\{(\hat m_i, l_i)\}$, with the understanding that the first summation goes through all node groups $\mathcal G_1, \cdots,\mathcal G_m$, and the second summation goes through each of the $r$ nodes in a given node group.  
} 
\begin{align}
&L^*QN\geq \eta_1\eta_2\sum_{\hat{m}=1}^{m}\sum_{\ell=1}^{r}\hat{m}^{\left(2r-2\ell\right)}(\hat{m}-1)^{(2\ell-1)} 
= \eta_1\eta_2\sum_{\hat{m}=1}^{m}\hat{m}^{2r-2}(\hat{m}-1)\sum_{\ell=0}^{r-1}\left( \frac{\hat{m}-1}{\hat{m}}\right)^{2\ell}
 \nonumber \\
&= \eta_1\eta_2\sum_{\hat{m}=1}^{m}(\hat{m}-1)\frac{\hat{m}^{2r}-( \hat{m}-1)^{2r}}{\hat{m}^2-(\hat{m}-1)^2} = \eta_1\eta_2\sum_{\hat{m}=1}^{m}\frac{\hat{m}-1}{2\hat{m}-1}(\hat{m}^{2r}-( \hat{m}-1)^{2r})  \nonumber  \\
&= \eta_1\eta_2\left(\sum_{\hat{m}=1}^{m}\frac{\hat{m}^{2r}-( \hat{m}-1)^{2r}}{2} - \sum_{\hat{m}=1}^{m}\frac{\hat{m}^{2r}-( \hat{m}-1)^{2r}}{4\hat{m}-2}\right) \nonumber\\ 
&= \eta_1\eta_2\left(\frac{m^{2r}}{2}- \sum_{\hat{m}=1}^{m}\frac{\hat{m}^{2r}}{4\hat{m}-2} + \sum_{\hat{m}=0}^{m-1}\frac{ \hat{m}^{2r}}{4\hat{m}+2} \right)
=\eta_1\eta_2\left(\frac{m^{2r}}{2}-\frac{m^{2r}}{4m-2} - \sum_{\hat{m}=1}^{m-1}\frac{ \hat{m}^{2r}}{4\hat{m}^2-1}\right).  
\label{eq: bound_prf_4} 
\end{align}
 By normalizing (\ref{eq: bound_prf_4}) by $QN=\eta_1\eta_2m^{2r}$, we obtain 
(\ref{eq:bound_homo_L*}).
\begin{comment}

Moreover, we can loosen the bound of (\ref{eq:bound_homo_L*})  to  find
\begin{align}
L^*& \geq \left(\frac{m^{2s}}{2}-\frac{m^{2s}}{4m-2} - \frac{1}{4}\sum_{\hat{m}=1}^{m-1}\hat{m}^{2s-2} \right)\geq \left(\frac{m^{2s}}{2}-\frac{m^{2s}}{4m-2} - \frac{1}{4}\sum_{\hat{m}=1}^{m-1}m^{2s-3}\hat{m} \right) \nonumber\\
&=\left(\frac{m^{2s}}{2}-\frac{m^{2s}}{4m-2} - \frac{m^{2s-3}}{4}\sum_{\hat{m}=1}^{m-1}\hat{m} \right)  = \left(\frac{m^{2s}}{2}-\frac{m^{2s}}{4m-2} - \frac{m^{2s-3}}{4}m\left(\frac{m-1}{2}\right) \right) \nonumber \\
& = \eta_1\eta_2\left(\frac{m^{2s}}{2}-\frac{m^{2s}}{4m-2} - \frac{m^{2s-1}}{8} + \frac{m^{2s-2}}{8}\right). \label{eq: bound_prf_2}
\end{align}
\end{comment}

Moreover, we can loosen the bound of (\ref{eq:bound_homo_L*})  to  find 
\begin{align}
L^*& \geq \frac{1}{2}-\frac{1}{4m-2} - \frac{1}{4} \left(\sum_{\hat{m}=1}^{m-1}\hat{m}^{2r-2}\right) m^{-2r} \geq 
\frac{1}{2}-\frac{1}{4m-2} - \frac{1}{8m} + \frac{1}{8m^2} 
\geq \frac{29}{96}.
\label{eq: Lstarm}
\end{align}
\begin{comment}
To obtain a bound for the communication load, we normalize (\ref{eq: bound_prf_2}) by $QN=\eta_1\eta_2m^{2r}$:
\be\label{eq: Lstarm}
L^* \geq \frac{1}{2}-\frac{1}{4m-2} - \frac{1}{8m} + \frac{1}{8m^2} \geq \frac{1}{3}-\frac{1}{32}=\frac{29}{96}.
\ee
\end{comment}
The last inequality in (\ref{eq: Lstarm}) follows from the left side of being an increasing function of $m$ when $m \ge 2$, and the minimum  is achieved at $m=2$. 
\begin{comment}
 Moreover, 
from (\ref{eq: eqL_sgt1}), we get
\begin{align}
L_{\rm c} &= 
\frac{1}{2}+\left(\frac{r}{K}\right)^r\left(  \frac{1}{4r-2}\left( \frac{K}{r}-1 \right)^r   - \frac{1}{2} \right)  \leq \frac{1}{2} + \frac{1}{4r-2} \leq \frac{2}{3}.
\label{eq:L_c_upper}
\end{align}
\end{comment}
Combining (\ref{eq: Lstarm}) with (\ref{eq:L_c_upper}), we obtain   (\ref{eq:Lc_upper}).
\begin{comment}
Putting it together we find
\be
\frac{1}{3}-\frac{1}{32} = \frac{29}{96}\leq L^* \leq L_{\rm c} \leq \frac{2}{3}
\ee
and
\be
\frac{ L_{\rm c}}{L^*} \leq \frac{64}{29} \approx 2.207.
\ee
\end{comment}
%
%
%
%
%

\section{Proof of Theorem \ref{th: conv}}
\label{sec: het_opt_pf}

 In the following, let $x_i = |\mathcal{K}_i|$ be the size of the $i$-th dimension of the hypercuboid. WLOG,  assume $x_1 \ge x_2 \ge \ldots \ge x_{s-1} \ge x_s$. 
First, We take a similar approach to Example 4 and the proof of  Theorem \ref{theorem: sg1_homo_bound} to derive $L_{P2}$.  With each node of the permutation we remove a layer of the hypercuboid. Through this process, the hypercuboid reduces in size as we disregard files available and functions assigned to nodes of the previous nodes of the permutation. In particular, we design the permutation such that the next node is aligned along the dimension with the largest remaining size (accounting for layers previously removed). For example, if after accounting for some nodes the remaining sizes of the dimensions are $\hat{x}_1,\ldots,\hat{x}_r$, we pick the next node from the set $\mathcal{K}_n$ such that $\hat{x}_n$ is the largest dimension. Then, we count the number of files of interests which is $\eta_1(\hat{x}_n-1)\prod_{j\neq n}\hat{x}_j$ and number of functions of interests $\eta_2\prod_{j\neq n}\hat{x}_j$
\begin{align}
L^*QN &\geq\eta_1\eta_2\sum_{i=1}^r\left(\left( \prod_{j=1}^{i-1}x_j^2\right)\sum_{j=1}^{x_i-x_{i+1}}\sum_{k=1}^{i}(x_i-j)^{2k-1}(x_i-j+1)^{2(i-k)} \right) \notag 
\end{align}
\begin{align}
&=\eta_1\eta_2\sum_{i=1}^r\left(\left( \prod_{j=1}^{i-1}x_j^2\right)\sum_{\hat{m}=x_{i+1}+1}^{x_i}\sum_{\ell=1}^{i}(\hat{m}-1)^{2\ell-1}\hat{m}^{2(i-\ell)} \right)
\label{eq:L_P2}
\end{align}
After scaling (\ref{eq:L_P2}) by $QN=\eta_1 \eta_2 X^2$, we obtain the desired expression for $L_{P2}$.

 Next, we derive 
  $L_{P1}$ using a different permutation that includes only the $x_1$ nodes aligned along the largest dimension.  { Note that since  nodes aligned along the same dimension collectively compute all functions,  the remaining nodes of the permutation are irrelevant.}
Each of the $x_1$ nodes computes $\eta_2\frac{X}{x_1}$ functions and there are $\eta_1X\frac{x_1-1}{x_1}$ files which are not available to it. For the first node of the permutation there are $\eta_1\eta_2X^2\left(\frac{x_1-1}{x_1^2}\right)$ IVs of interest using the bound of Lemma \ref{theorem: bound}. Since nodes aligned along the same dimension do not have any assigned functions in common,  the number functions of interest remains the same for the following nodes. However, the number of files of interest decreases by $\eta_1\frac{X}{x_1}$ for each following node of the permutation. Since nodes aligned along the same dimension do not have any available files in common, the number of files of interest decreases by the same amount with each node in the permutation. Thus,
\begin{align}
  L^*QN &\geq \sum_{i=1}^{x_1}\eta_1\eta_2\left(\frac{X}{x_1}\right)\left(X\frac{x_1-1}{x_1} - (i-1) \cdot\frac{X}{x_1}\right) 
  = \eta_1\eta_2\frac{X^2(x_1-1)}{2x_1}.
  \label{eq:L_P1}
\end{align}
By combining (\ref{eq:L_P1}) and (\ref{eq:L_c_upper}), we obtain (\ref{eq:constant_het}). 

\if{0}
\subsection{Optimality Example}

\mj{[MJ: include in the ArXiv version]}
In the following, we provide an example of finding a lower bound on the achievable communication load given the file placement and function assignment of the previous $3$-dimensional example. To do this we use Theorem \ref{theorem: bound} which provides a lower bound on the entropy of a set of messages $X_\mathcal{K}$ transmitted among the nodes to satisfy all node requests for successful MapReduce execution.

We derive a bound on the optimal communication load, $L^*$, for the 3-dimensional example by picking a permutation of the nodes and using Theorem \ref{theorem: bound}. The permutation we use here is $(1,7,6,2,8,5,3,9,4)$. Most importantly, to achieve a tighter bound, the permutation contains $3$ sequential sets of nodes such that each set contains one node aligned along each dimension of the cube. In order to calculate the necessary terms of (\ref{eq: bound_eq1}), we can simply count the number of files in $\mathcal{M}_{\mathcal{K}}\setminus \mathcal{M}_{\{k_1,\ldots , k_{i} \}}$ and the number of functions in $\mathcal{W}_{k_i}\setminus \mathcal{W}_{\{k_1,\ldots , k_{i-1} \}}$ as their product represents the number of intermediate values of interest. Moreover, since the intermediate values are independent and of size $T$ bits, we can conclude that
\begin{align}
H&\left(V_{\mathcal{W}_{k_i},:}|V_{:,\mathcal{M}_{k_i}},Y_{\{k_1,\ldots, k_{i-1} \}}\right)  \nonumber \\
&=T\cdot \left|\mathcal{M}_{\mathcal{K}}\setminus \mathcal{M}_{\{k_1,\ldots , k_{i} \}}\right| \cdot \left|\mathcal{W}_{k_i}\setminus \mathcal{W}_{\{k_1,\ldots , k_{i-1} \}}\right|.
\end{align}
In Fig.~\ref{fig: 3d_proof_fig}, we have highlighted the lattice points representing the sets of files and functions which are used to obtain the bound. The lattice points representing the files are highlighted in red and the lattice points representing the functions are highlighted in green. First, we simply consider every function assigned to node $1$ and every file not available to node $1$, where node $1$ is the first node in the permutation. This is shown in figure Fig.~\ref{fig: 3d_proof_fig}(a). We see that
\be
H\left(V_{\mathcal{W}_{1},:}|V_{:,\mathcal{M}_{1}}\right) = T\cdot 18 \cdot 9 = 162T
\ee
since in this case each lattice point only represents $1$ file and $1$ function ($\eta_1=\eta_2=1$).

\begin{figure}
\centering
\centering \includegraphics[width=9cm]{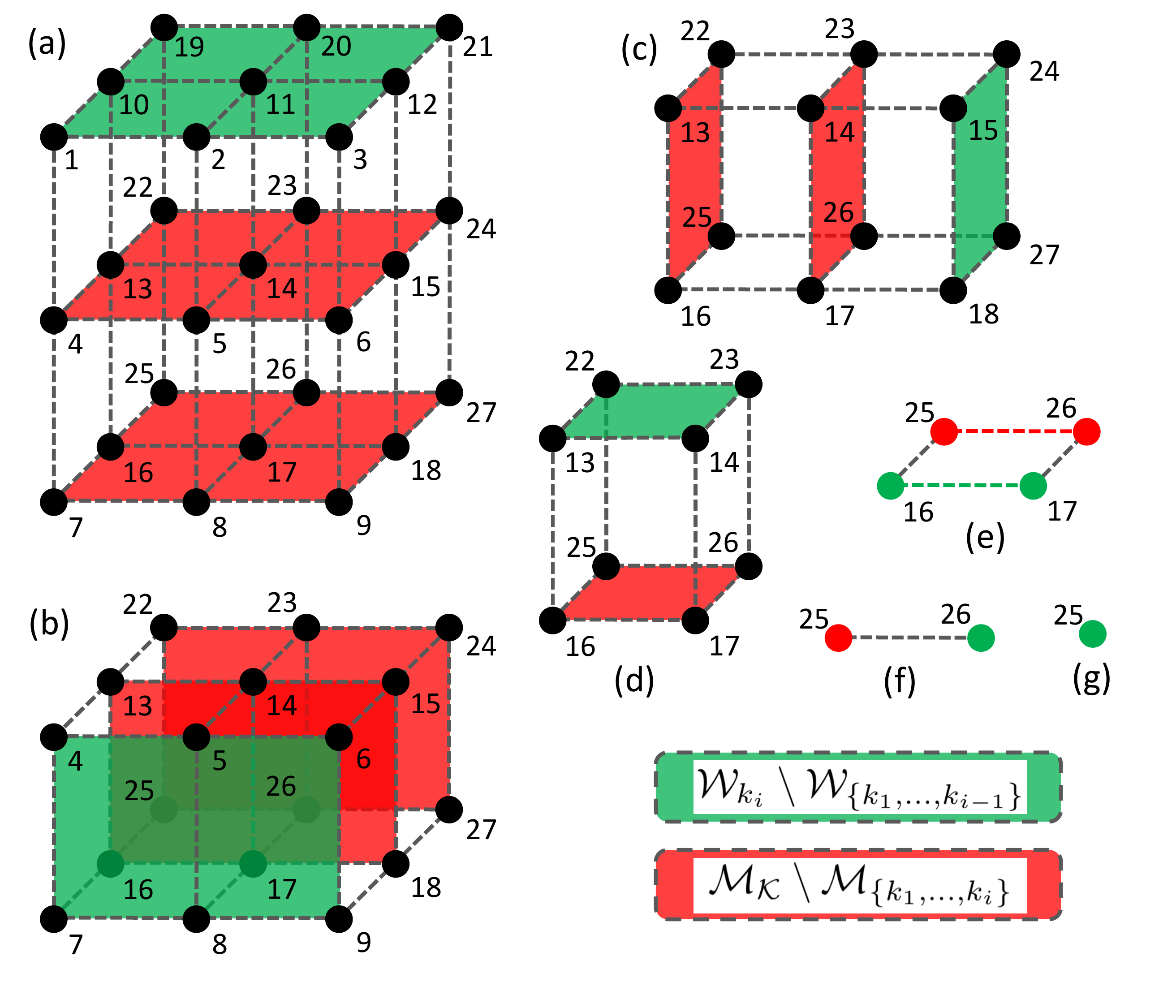}
\vspace{-0.5cm}
\caption{~\small A representation of the optimality example of Section \ref{sec: opt_exp}. In a) through g) functions of interest are highlighted green and files of interest are highlighted red. Between each step, the lattice shrinks in one dimension by one unit. }
\label{fig: 3d_proof_fig}
\vspace{-0.4cm}
\end{figure}

Similarly, for node $7$, we are interested in functions it computes and files it does not have locally available, except this time we do not count files available to node $1$ or functions assigned to node $1$. Fig.~\ref{fig: 3d_proof_fig}(b) shows the files and functions we are counting. Note that, we disregard the top layer of the cube which represents the files and functions assigned to node $1$. We see that
\be
H\left(V_{\mathcal{W}_{7},:}|V_{:,\mathcal{M}_{7}},Y_{\{1 \}}\right) = T\cdot 6 \cdot 12 = 72T.
\ee
By continuing this process, from Fig.~\ref{fig: 3d_proof_fig}(c-f), we see that
\begin{align}
H\left(V_{\mathcal{W}_{6},:}|V_{:,\mathcal{M}_{6}},Y_{\{1,7 \}}\right) &= 48T \\
H\left(V_{\mathcal{W}_{2},:}|V_{:,\mathcal{M}_{2}},Y_{\{1,7,6 \}}\right) &= 16T \\
H\left(V_{\mathcal{W}_{8},:}|V_{:,\mathcal{M}_{8}},Y_{\{1,7,6,2 \}}\right) &= 4T \\
H\left(V_{\mathcal{W}_{5},:}|V_{:,\mathcal{M}_{5}},Y_{\{1,7,6,2,8 \}}\right) &= T.
\end{align}

Finally, enough lattice points are removed from the cube and only $1$ lattice point remains as shown in Fig.~\ref{fig: 3d_proof_fig}(g) which represents the function assigned to node $3$. However, we see there are no lattice points representing files node $3$ does not have locally available. This occurs because the other two nodes aligned along the same dimension, $1$ and $2$, as node $3$ have already been accounted for, and those two nodes have collectively have all the files node $3$ does not have. Therefore,
\be
H\left(V_{\mathcal{W}_{3},:}|V_{:,\mathcal{M}_{3}},Y_{\{1,7,6,2,8,5 \}}\right) = 0.
\ee
Similarly, for the remaining two nodes of the permutation of $\mathcal{K}$, nodes $9$ and $4$, there are no remaining files that are not locally available to them. In fact, there are also no functions assigned to nodes $9$ and $4$ which have not already been accounted for. Therefore,
\begin{align}
H\left(V_{\mathcal{W}_{9},:}|V_{:,\mathcal{M}_{9}},Y_{\{1,7,6,2,8,5,3 \}}\right) &= 0 \\
H\left(V_{\mathcal{W}_{4},:}|V_{:,\mathcal{M}_{4}},Y_{\{1,7,6,2,8,5,3,9 \}}\right) &= 0.
\end{align}

By taking the sum of (\ref{eq: bound_eq1}), we directly compute the bound of (\ref{eq: bound_prf_4}) and find that
\begin{align}
LNQT &\geq H(X_\mathcal{K}) \geq 303T \\
L &\geq \frac{303T}{QNT} = \frac{303}{729} \approx 0.416.
\end{align}
\fi

\section{Proof of Corollary \ref{corollary:optimality}}
\label{sec: pf_homo_compare}


For (a), given that $r=s=2$, we obtain $L_1=\frac{2(K-2)}{3(K-1)}$ from (\ref{eq: LiResult}), and $L_c=\frac{2}{3}-\frac{1}{2X} \frac{K+2}{3}$ from  (\ref{eq: eqL_sgt1}) using $|\mathcal{K}_1|+|\mathcal{K}_2|=K$
and  $|\mathcal{K}_1| \cdot |\mathcal{K}_2|=X$. Since $L_c$ is the largest when $X$ is maximized to be $X=(\frac{K}{2})^2$ (corresponding to the homogeneous network), we have $L_c \le \frac{2(K+1)(K-2)}{3K^2} < L_1.$

Next, for (b), when $r=s \leq \frac{K}{2}$ such that there exists an achievable hypercuboid design, then ${\min \{ r+s, K \} = r + s = 2r}$. By only considering the last term of $L_1(r,r)$ in (\ref{eq: LiResult}) we derive the following lower bound.
\begin{align}
L_1(r,r) &> \frac{2r{K \choose 2r}{2r-2 \choose r-1}{ r \choose r}}{r{K \choose r}^2} 
= \frac{r}{2r-1}\cdot \frac{(K-r)(K-r-1) \cdots (K-2r+1)}{K(K-1) \cdots (K-r+1)} \nonumber \\
& > \frac{r}{2r-1}\left( 1-\frac{r}{K-r+1}\right)^r = \frac{r}{2r-1}\cdot (1 + o(1)) \label{eq: L1_low_bound}
\end{align}
 Next, we derive an upper bound on $L_c$.
 For a given $r$ and $K$, let $|\mathcal{K}_1|=\cdots =|\mathcal{K}_{r-1}|=2$ and $|\mathcal{K}_r|=K-2(r-1)$. Then by (\ref{eq: eqL_sgt1})
  \be \label{eq: het_low_bound}
  L_{\rm c} = \frac{X-1}{2X} + \frac{K-2r+1}{2^{r-1}(K-2r+2)(4r-2)} < 
  \frac{r}{2r-1}\left( 1-\frac{1}{2r} + \frac{1}{r2^r} \right).
  \ee
  Then, combining (\ref{eq: L1_low_bound}) and (\ref{eq: het_low_bound}) we find $L_{\rm c}<L_1$ if
  \begin{align}\label{eq: condition_compare}
      K > r - 1 +\frac{r}{1-\left( 1-\frac{1}{2r} + \frac{1}{r2^r}  \right)^{1/r}}. \
  \end{align}
  We now aim to find an upper bound on the RHS of (\ref{eq: condition_compare}). It can be shown that if $r\geq 6$ then $\left( 1-\frac{1}{2r} + \frac{1}{r2^r}  \right)^{1/r} \leq \exp (-\frac{1}{2r^2})$ and $1-\exp (-\frac{1}{2r^2})>\frac{1}{4r^2}$. Substituting this into (\ref{eq: condition_compare}), we find $L_1<L_{\rm c}$ if $r\geq6$ and $K>r-1+4r^3$ which proves (b).

From (\ref{eq: L1_low_bound}), if $r=\Theta(1)$ then $L_1(r,r)\geq\frac{r}{2r-1}+o(1)$, and alternatively, if $r = \Omega(1)$ and $r=o(K)$ then $L_1(r,r)\geq\frac{1}{2}+o(1)$. From (\ref{eq:L_c_upper}), $L_{\rm c} <  \frac{r}{2r-1}$.  Therefore, with the given assumptions that $r\geq 1$ and $r = o(K)$, we find that
  $ \frac{L_{\rm c}(r)}{L_1(r,r)} \le 1 + o(1)$ which proves (c). 
  

\begin{comment}
\section{Proof of Theorem \ref{theorem: heterogeneous ub}}
\label{sec: het_sgt1_pf}
From (\ref{eq: het_sgt1_bound})  we see that $L_{\rm c} \le \frac{r}{2r-1}$. Moreover, it was shown in (\ref{eq: L1_low_bound}) that $L_1 > \frac{r}{2r-1}\cdot (1 + o(1))$ when $r=o(K)$. Therefore,
\be
\lim_{r \rightarrow \infty}\frac{L_{\rm c}(r)}{L_1(r,r)} \leq 1 + o(1).
\ee
\end{comment}

\bibliographystyle{IEEEbib}
\bibliography{references_d2d}

\end{document}